\documentclass[openacc]{rstransa}

\titlehead{Research}

\begin{document}

\title{Dual-phase xenon time projection chambers for rare-event searches}

\author{
Laura Baudis}

\address{Physik-Institut, University of Z\"urich,  Winterthurerstrasse 190, 8057  Z\"urich, Switzerland}

\subject{Xenon TPC, dark matter, neutrinos, weak decays}


\corres{Insert corresponding author name\\
\email{laura.baudis@uzh.ch}}

\begin{abstract}
In the past decade, dual-phase xenon time projection chambers (Xe-TPCs) have emerged as some of the most powerful detectors in the fields of astroparticle physics and rare-event searches. Developed primarily towards the direct detection of dark matter particles, experiments presently operating deep underground have reached target masses at the multi-tonne scale, energy thresholds around 1\,keV and radioactivity-induced background rates similar to those from solar neutrinos. These unique properties, together with demonstrated stable operation over several years, allow for the exploration of new territory via high-sensitivity searches for a plethora of ultra-rare interactions. These include searches for particle dark matter, for second order weak decays, and the observation of astrophysical neutrinos. We first review some properties of xenon as a radiation detection medium and the operation principles of dual-phase Xe-TPCs together with their energy calibration and resolution. We then discuss the status of currently running experiments and of proposed next-generation projects, describing some of the technological challenges. We end by looking at their sensitivity to dark matter candidates, to second order weak decays and to solar and supernova neutrinos. Experiments based on dual-phase Xe-TPCs are difficult, and, like all good experiments, they are constantly pushed to their limits.  Together with many other endeavours in astroparticle physics and cosmology they will continue to push at the borders of the unknown, hopefully to reveal profound new knowledge about our cosmos.

\end{abstract}


\maketitle

\section{Introduction}
The idea of a Time Projection Chamber (TPC), "born on a waning winter afternoon" in February 1974~\cite{Nygren:2018sjx} was first realised and successfully applied in particle physics experiments at accelerators. In the last decades, the concept evolved and was tailored to a large range of applications, most notably to experiments in astroparticle and nuclear physics physics searching for rare interactions deep underground. Present and future experiments designed to observe  interactions of dark matter and other exotic particles, second order weak nuclear decays, as well as neutrinos from a variety sources are based on TPCs operated with pure noble elements, either in gaseous or liquid form. As in the original idea by David Nygren~\cite{Nygren:1976fe}, modern TPCs capture the $x-y$ information of events, along with their drift time, allowing for a three-dimensional position reconstruction if the start time is known. 

The two most common noble fluids employed as detector media in present TPCs for rare event searches are argon and xenon.  Here we will focus on dual-phase (liquid and gas) xenon TPCs (Xe-TPCs). For general reviews of experiments based on noble elements, we refer the reader to \cite{Aprile:2009dv,Chepel:2012sj,Gonzalez-Diaz:2017gxo}. Historical introductions, the physics and technological details, as well as applications of noble gas and electron emission detectors are detailed in \cite{Aprile:2008bga,Bolozdynya:2010zz}.

This article is structured as follows. In Section~\ref{sec:properties} the properties of xenon as a radiation detection medium will be briefly reviewed.  The working principle of two-phase Xe-TPCs and their three-dimensional imaging capabilities are introduced in Section~\ref{sec:tpc}, while their energy calibration and resolution is discussed in Section~\ref{sec:energy}. Section~\ref{sec:experiments} looks at past, current and future detectors, and at some technological challenges.  Section~\ref{sec:dm} and~\ref{sec:doubleweak}  will discuss the search for dark matter and other exotic particles, and for second order weak decays, respectively. Section~\ref{sec:neutrinos} will scrutinise  the potential of dual-phase Xe-TPCs to detect neutrinos from the Sun and from core-collapse supernovae, while Section~\ref{sec:conclusions} will conclude the brief review.

 \section{Properties of xenon as radiation detection medium}
 \label{sec:properties}
 
Xenon is an excellent scintillator and good ioniser in response to the passage of radiation. In a TPC, the simultaneous detection of ionisation and scintillation allows for the identification of the primary particle interacting in the medium based on the linear energy transfer, $dE/dx$, and for the determination of the three-dimensional position of an interaction with sub-mm (in the $z$-coordinate) to mm (in the $x-y$-coordinate) precision. Table \ref{tab:properties} lists some of the physical properties  of xenon, relevant for practical aspects of a detector. The high atomic mass number and high liquid density allows for compact, large and homogeneous detector geometries with efficient self-shielding against external radiation, given that the cross sections for the photoelectric effect, Compton scattering and pair production scale as $Z^5/E_{\gamma}^{7/2}$, $Z/E_{\gamma}$ and $Z^2\ln{(2 E_{\gamma})}$, respectively, for incoming X-rays and gammas with energy $E_{\gamma}$.  The radioactive isotopes $^{124}$Xe, $^{126}$Xe, $^{134}$Xe and $^{136}$Xe have very long half-lives, and their second-order weak decay modes are subject to investigation, as we will show in Section~\ref{sec:doubleweak}.
 
\begin{table}[!h]
\caption{Physical properties, volume fraction in the atmosphere, radioactive isotopes of the noble element xenon.}
\centering
\label{tab:properties}
\begin{tabular}{lc}
\hline
{Property [unit]} & {Xe}   \\ 
\hline
Atomic number & 54    \\
Mean atomic mass [g/mol] & 131.29  \\
Boiling point $T_b$ at 1 atm [K] 	& 165.0   \\
Melting point $T_m$ at 1 atm [K]	& 161.4  \\
Gas density at 1 atm \& 298 K [g/l] 	& 5.40  \\
Gas density at 1 atm \& $T_b$ [g/l] 	& 9.99   \\
Liquid density at $T_b$ [g/cm$^3$]	 & 2.94   \\
Volume ratio & 	526  \\ 
Dielectric constant of liquid &	1.95  \\
Volume fraction in Earth's atmosphere [ppm] &	0.087 \\
Isotopes with spin, abundance [\%] &	$^{129}$Xe, 26.44;\,\,  $^{131}$Xe, 21.18 \\
Radioactive isotopes, &	$^{136}$Xe, 8.87;\,\,  2.2$\times$10$^{21}$ ~\cite{Albert:2013gpz,KamLAND-Zen:2019imh} \\
abundance  [\%]  and  T$_{1/2}$ [y] &	$^{124}$Xe, 0.095;\,\, 1.1$\times$10$^{22}$ ~\cite{XENON:2019dti,XENON:2022evz} \\
 &	$^{134}$Xe, 10.4;\,\, $>$8.7$\times$10$^{20}$ (90\% C.L.)~\cite{EXO-200:2017vqi} \\
  &	$^{126}$Xe, 0.089;\, $>$1.9$\times$10$^{22}$ (90\% C.L.)~\cite{XMASS:2018txy} \\
\hline
\end{tabular}
\vspace*{-4pt}
\end{table}

In a xenon detector, the energy loss of an incident particle is shared between ionisation, excitation  and sub-excitation electrons liberated in the ionisation process. The average energy loss in ionisation is slightly larger than the ionisation potential, as it includes multiple ionisation processes.   The energy $E_0$ transferred to the medium, for light particles such as electrons, is~\cite{Platzman:1961}:
\begin{equation}
E_0 = N_i E_i + N_{ex} E_{ex} + N_i \epsilon, 
\label{eq:energy_transfer}
\end{equation}
where $N_i$ and $N_{ex}$ are the mean number of ionised and excited atoms, $E_i$ and $E_{ex}$ are the mean energies  to ionise or excite the atoms and $\epsilon$ is the average kinetic energy of sub-excitation electrons, the energy of which goes into heat. For liquid xenon, the value of  $\epsilon$  is in the range $4.65-5.25$\,eV.  In its condensed phase, xenon exhibits a band structure of electronic states, and if we divide all terms in Eq.~\ref{eq:energy_transfer} by the band gap energy $E_g$ and uses the $W_i$-value, defined as the energy required to produce one electron-ion pair, $W_i=E_0/N_i$, we obtain:

\begin{equation}
\frac{W_i}{E_g} = \frac{E_i}{E_g} + \frac{N_{ex}}{N_i} \times \frac{E_{ex}}{E_g}  + \frac{\epsilon}{E_g}, 
\label{eq:energy_normalised}
\end{equation}
The band gap energy is 9.22\,eV and 9.28\,eV for liquid and solid xenon, respectively, while the ratio $\alpha = N_{ex}/{N_i}$ is between 0.06 and 0.2. For nuclear recoils, the ratio $N_{ex}/{N_i}$ is 1, and thus much larger than for electronic recoils.  This property is employed in dark matter searches via nuclear recoils, discussed in Section~\ref{sec:dm}. The ratio $W_i/E_g$ has been calculated as 1.65, in good agreement with measurements, which yield 1.6~\cite{Aprile:2009dv}.

Scintillation arises from excited xenon atoms $\mathrm{Xe}^*$ (excitons) and from ions $\mathrm{Xe}^+$, as follows:
\begin{enumerate}
\item 
\[{\mathrm {Xe}}^*+{\mathrm {Xe}}+{\mathrm  {Xe}}\rightarrow {\mathrm  {Xe}}^{*}_{2}+{\mathrm  {Xe}}\]
\[{\mathrm  {Xe}}^{*}_{2}\rightarrow 2{\mathrm  {Xe}}+{\mathrm h\nu}\] 
\item 
\[{\mathrm  {Xe}}^{+}+{\mathrm  {Xe}} \rightarrow {\mathrm  {Xe}}^{+}_{2}\]
\[{\mathrm  {Xe}}^{+}_{2}+\mathrm{e}^-\rightarrow {\mathrm  {Xe}}^{**}+{\mathrm  {Xe}}\]
\[{\mathrm  {Xe}}^{**}\rightarrow {\mathrm  {Xe}}^*+{\mathrm {heat}}\]
\[{\mathrm  {Xe}}^*+{\mathrm  {Xe}}+{\mathrm  {Xe}}\rightarrow {\mathrm  {Xe}}^{*}_{2}+{\mathrm {Xe}}\]
\[{\mathrm  {Xe}}^{*}_{2}\rightarrow 2{\mathrm  {Xe}}+{\mathrm h\nu}\]
\end{enumerate}
where h$\nu$ denotes the emitted vacuum-ultraviolet (VUV) photon with wavelength peaked around 175-178\,nm.  $\mathrm{Xe}^{**}\rightarrow \mathrm{Xe}^*$ + heat corresponds to a non-radiative transition. The excited dimer $\mathrm{Xe}^{*}_{2}$ (excimer), at its lowest excited level, is de-excited to the dissociative ground state by the emission of a single VUV photon. 
This comes from the large energy gap between the lowest excitation and the ground level, forbidding other decay channels such as non-radiative transitions.  The scintillation light from pure liquid  xenon has two decay components due to de-excitation of spin singlet ($^{1}\Sigma_u^+$) and spin triplet ($^{3}\Sigma_u^+$) states of the excited dimer ${\mathrm Xe}^{*}_{2}\rightarrow 2{\mathrm Xe}+{\mathrm h\nu}$ to the ground state ($^{1}\Sigma_g^+$).  The singlet and triplet states refer to the total spin quantum number (s=0 or s=1) of the excited Rydberg electron and the angular momentum due to the molecular orbit, with the shorter and longer decay shapes being produced by the de-excitation of s=0 states and  s=1 states, respectively. The differences of pulse decay shape between different type of particle interactions can be in principle be used to  discriminate electronic from nuclear recoils. In practice, this is difficult due to the small time separation, $(2.2\pm0.3)$\,ns versus $(27\pm1)$\,ns, of the two components.

If we denote $W_{ph}$ as the mean energy required for the production of a single photon, we can express it as:
\begin{equation}
W_{ph} = \frac{E_0}{N_{ex} + N_i} =  \frac{W_i}{1 + N_{ex}/N_i}=  \frac{W_i}{1 + \alpha},
\label{eq:wvalue_photon}
\end{equation}
where $N_{ex}, N_i$ are the produced number of excitons and electron-ion pairs, respectively, and  we assume for simplicity that the efficiencies for an exciton or a recombining electron-ion pair to create a detectable photon are unity:
\begin{equation}
N_{ph} =  N_{ex} + r N_i,
\end{equation}
with $r$ being the recombination fraction.  If an electric field is applied, one can measure the electrons which do not recombine, with the amount of extracted charge defined as:
\begin{equation}
N_q =  (1-r) N_i
\label{eq:extractedcharge}
\end{equation}
Using  above equations, the recombination independent sum can be defined, following~\cite{DahlThesis}, as:
\begin{equation}
E_0=  (N_q + N_{ph}) \times W_{ph} 
\label{eq:sum}
\end{equation}
The recombination independent energy required to produce a single detectable quantum, $N_q$ or $N_{ph}$, is often called the W-value, and  $ W_{ph}  = W$ is used in the following. This assumes that  each recombining electron-ion pair produces an exciton, which leads to a photon. The widely-adopted W-value in liquid xenon has been measured as $W=(13.7\pm0.2$)\,eV with a small, 30\,g LXe detector using 122\,keV $\gamma$-rays from an external $^{57}$Co source~\cite{DahlThesis}. Recently, a lower value  of $W=11.5^{+ 0.2}_{- 0.3}$ (syst.)\,eV was measured with a small TPC using internal sources at energies between 2.8\,keV and 42\,keV~\cite{Baudis:2021dsq}, which is consistent with the value measured at the MeV-scale, $W=11.5\pm0.5 \mathrm{(syst.)} \pm 0.1 \mathrm{(stat.)}$\,eV with a larger detector~\cite{Anton:2019hnw}. 
\begin{figure}[!h]
\centering
\includegraphics[width=11 cm]{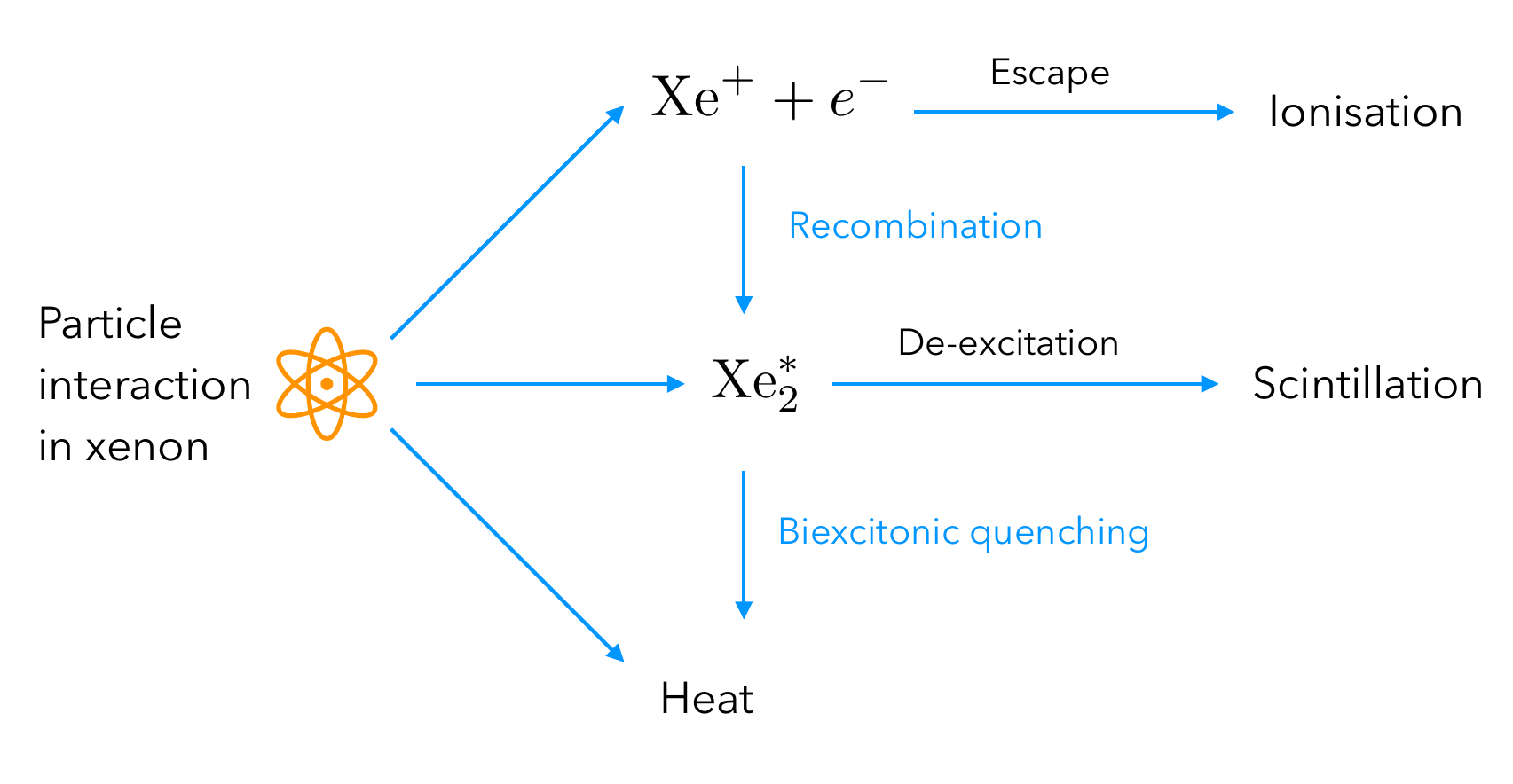}
\caption{After a particle interacts in a xenon medium, various processes lead to ionisation, scintillation and heat. Only scintillation and ionisation (via electroluminiscence) are observed in a two-phase Xe-TPC. Biexcitonic quenching~\cite{Hitachi:2005ti} is relevant for interaction with high ionisation density.}
\label{fig:intxenon}
\end{figure}   

The partition between excitation and ionisation depends on the density of the electron-ion pairs produced along the track of a particle, and is thus different for nuclear and electronic recoils. The recombination fraction $r$ depends on the applied electric field, as well as on the ionisation density in the track. In particular, for nuclear recoils as generated by interactions of fast neutrons or hypothetical Weakly Interacting Massive Particles (WIMPs), an energy-dependent quenching is introduced via the Lindhard factor $\mathcal{L}$~\cite{Lindhard:1963} :
\begin{equation}
E_0=  \mathcal{L}^{-1}  (N_q + N_{ph}) W, 
\label{eq:sum_lindhard}
\end{equation}
with $\mathcal{L}$ being around $0.15-0.2$ at nuclear recoil energies in the range 3-100\,keV~\cite{Sorensen:2011bd}.  A schematic view of the signal production after a particle interaction in xenon is shown in Figure~\ref{fig:intxenon}.
 
\section{Principles of dual-phase xenon TPCs}
\label{sec:tpc}
 
Xenon TPCs in astroparticle physics and rare-event searches use either high-pressure gas, or a liquid phase, or liquid and gas phase (also called two-phase or dual-phase) as detection medium, and here we focus on the latter.  Table \ref{tab:propertiesTPC} lists some properties relevant for building a TPC.
\begin{table}[!h]
\caption{Properties of liquid xenon as a particle detection medium, relevant for building a TPC.}
\centering
\label{tab:propertiesTPC}
\begin{tabular}{lc}
\hline
{Property [unit]} & {Value}  \\ 
\hline
Scintillation light yield (at 122\,keV)~\cite{Lenardo:2014cva} & 63 photons/keV    \\
Wavelength (peak centred at) & 175-178\,nm \\
Decay time constants (s=0, s=1) & 2.2\,ns, 27\,ns \\
Refractive index  & 1.69 \\
Electron mobility~\cite{Baudis:2023ywo}	& 0.29\,mm$^2$/($\mu$s\,V) ($<$100\,V/cm)  \\
	& 0.01\,mm$^2$/($\mu$s\,V) ($>$100\,V/cm)  \\
\hline
\end{tabular}
\vspace*{-4pt}
\end{table}

\begin{figure}[!h]
\centering
\includegraphics[width=8 cm]{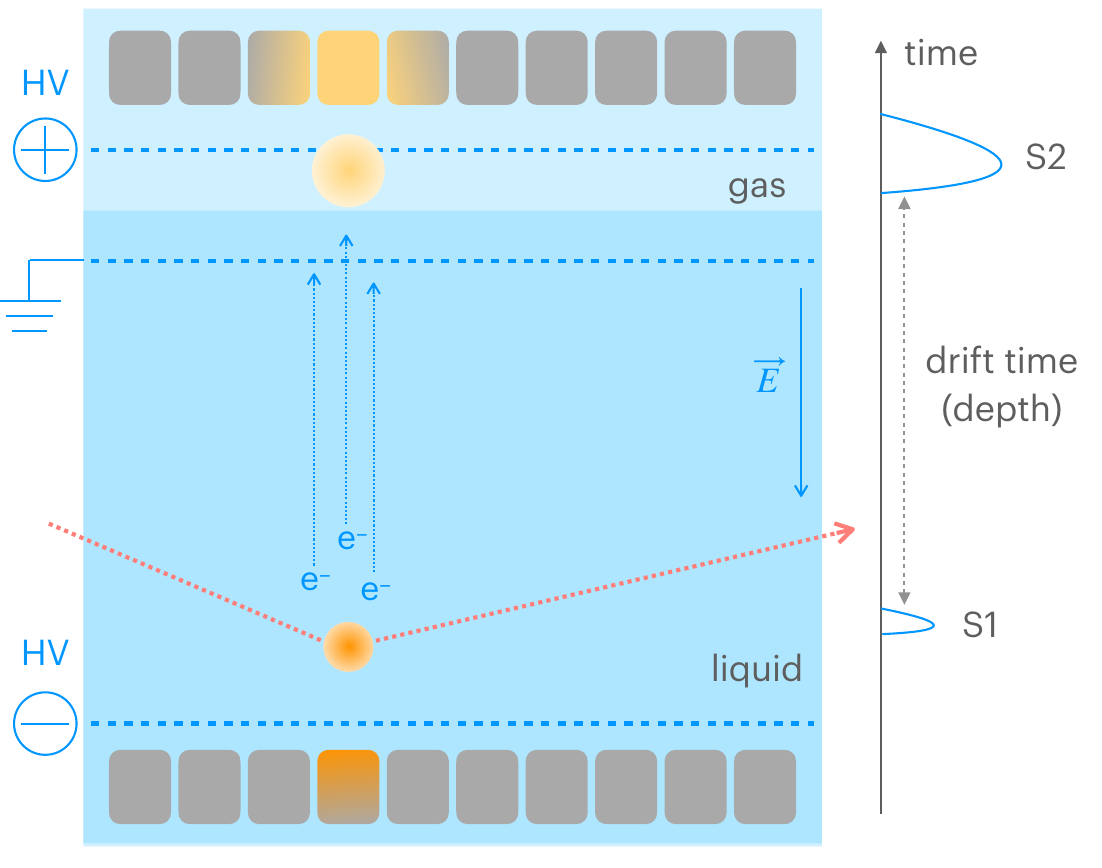}
\caption{The operation principle of a two-phase xenon TPC. A particle interaction in liquid xenon gives rise to a prompt scintillation signal (S1) and a delayed, amplified proportional scintillation signal (S2). The latter is caused by ionisation electrons, which are drifted in a homogeneous electric field (of a few 100\,V/cm) and extracted into the gas phase above the liquid with a higher electric field, typically 10\,kV/cm. The drift field is produced between the cathode at negative potential and a grounded gate grid in the liquid, while the extraction field is obtained by means of the anode placed above the gate in the gas phase. Both S1 and S2 signals are observed with photosensor arrays placed on the bottom and top of the TPC.}
\label{fig:schematictpc}
\end{figure}

The operation principle of a two-phase TPC is easy to grasp, and it is shown schematically in Figure~\ref{fig:schematictpc}.  An interaction within the  active volume of a detector will create ionisation electrons and prompt scintillation photons. The prompt scintillation signal (S1) is detected with two arrays of photosensors, one in the liquid phase on the bottom and one in the gas phase at the top. The electrons drift in the pure liquid under the influence of an external electric field, are then accelerated by a stronger field and extracted into the vapour phase above the liquid, where they generate proportional scintillation, or electroluminiscence.  The delayed, proportional scintillation signal (S2) is observed by the same photosensor arrays. The array immersed in the liquid collects the majority of the prompt signal, which is totally reflected at the liquid-gas interface. The ratio of the two signals is different for nuclear recoils, such as from fast neutron interactions or  hypothetical {\small WIMPs} and electronic recoils produced by $\beta$ and $\gamma$-rays. This provides the basis for background discrimination in dark matter detectors.  Since electron diffusion in the ultra-pure liquid is small (albeit non-negligible, see Section~\ref{subsec:challenges}), the proportional scintillation photons carry the $x-y$ information of the interaction site. With the $z-$information from the drift time measurement, the {\small TPC} yields  a three-dimensional event localisation, enabling fiducial volume selections and differentiation between single- and multiple-scatters in the active volume.  

 \section{Energy calibration and resolution of xenon TPCs}
 \label{sec:energy}

\begin{figure}[!h]
\includegraphics[width= 6.1 cm]{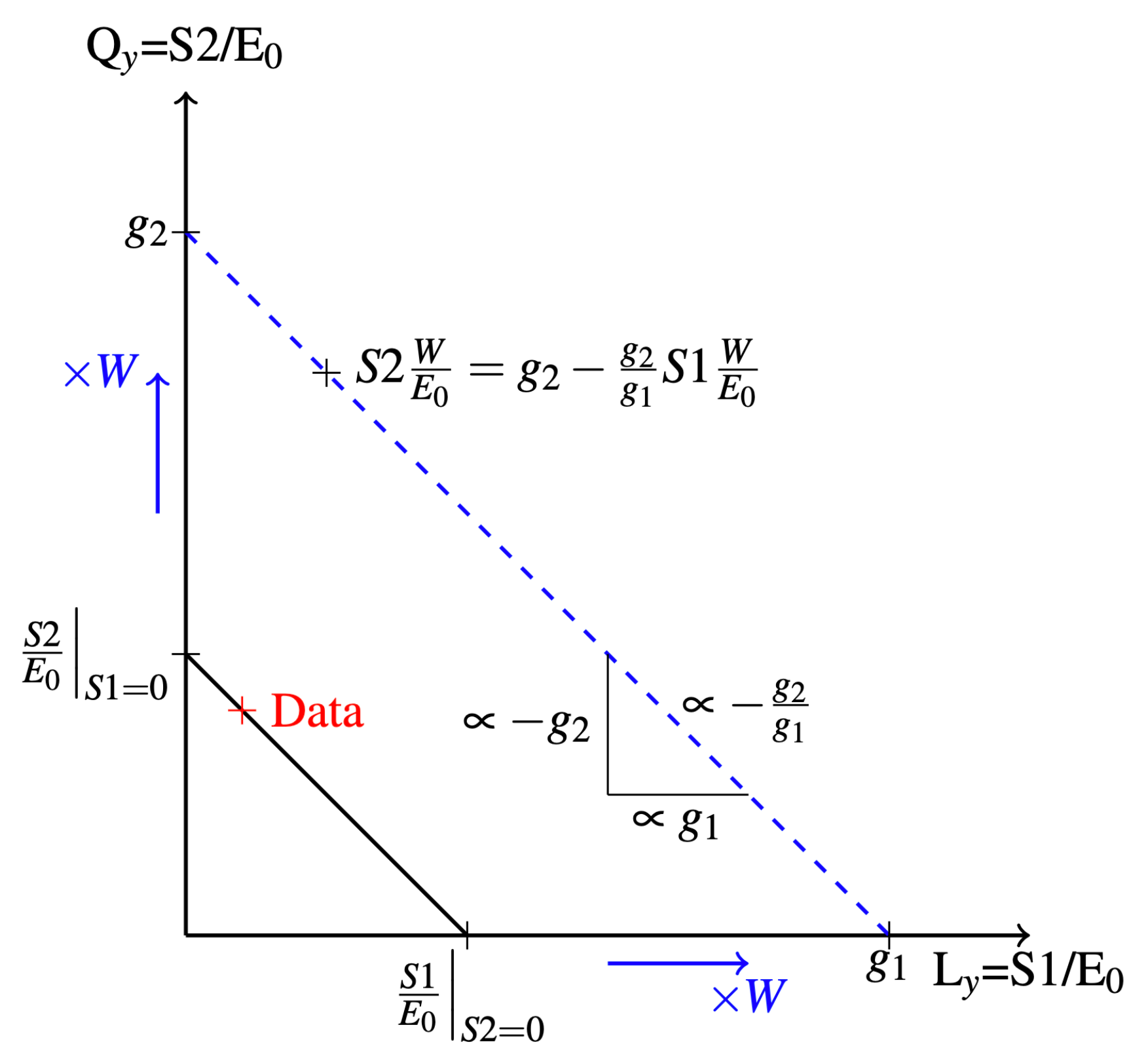}
\includegraphics[width= 7.8 cm]{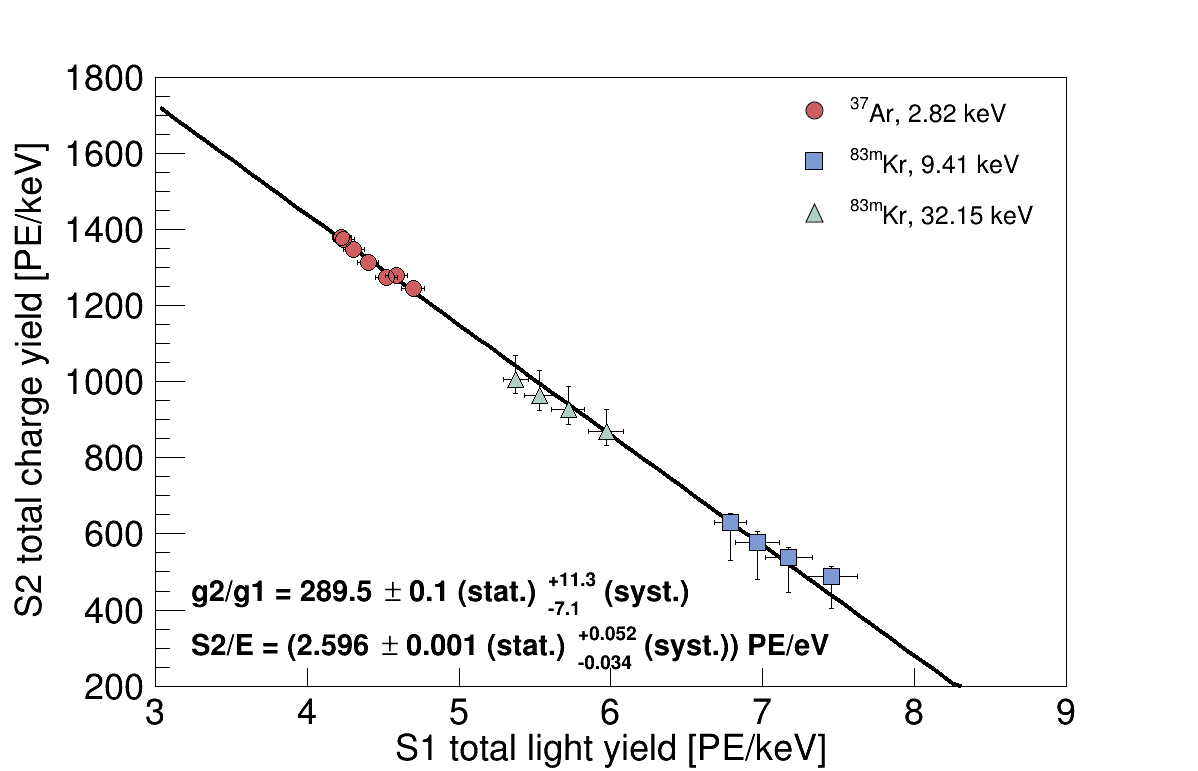}
\caption{(Left): Schematic of a Doke plot in charge (Q$_y$) versus light (L$_y$) yield for an interaction energy E$_0$. Measurements of the charge and light yield at different electric drift fields or interaction energies yield data along the solid line. The dashed line, showing the relation to the gain parameters $g_1$ and $g_2$, is obtained by scaling with the $W$-value. (Right): Normalised scintillation and ionisation signals at low energies for different drift fields in the range 80-968\,V/cm for a small TPC, Xurich. Figures from~\cite{Baudis:2021dsq}.}
\label{fig:doke}
\end{figure}

An advantage of two-phase Xe-TPCs is their relatively good energy resolution, achieved by taking a  linear combination of the two anti-correlated signals, S1 and S2.  To calibrate the energy scale, mono-energetic lines from external (e.g., $^{57}$Co, $^{137}$Cs, $^{228}$Th, etc) and internal ($^{83m}$Kr, $^{37}$Ar) calibration sources are used. Additional signals are provided by neutron-activated xenon ($^{129m}$Xe, $^{131m}$Xe) and the radioactivity of detector components (e.g., $^{60}$Co, $^{208}$Tl). The mean detected S1 and S2 signals per produced photon and ionisation electron are denoted as g$_1$ and g$_2$, respectively: $g_1 = S1/N_{ph}$,  $g_2 = S2/N_q$, also called the total photon detection efficiency and the charge amplification factor. We can then express equation~\ref{eq:sum} as:
\begin{equation}
E_0=  (N_{ph}  + N_q ) \times W = \left( \frac{S1}{g_1} + \frac{S2}{g_2} \right) \times W
\label{eq:sumgain}
\end{equation}
Rearranging this equation yields:
\begin{equation}
\frac{S2}{E_0}=  \frac{g_2}{W} - \frac{g_2}{g_1} \frac{S1}{E_0}
\label{eq:sumgaindoke}
\end{equation}
Since W and E$_0$ (for a given mono-energetic signal) are known, we can determine g$_1$ and g$_2$ from the measured S1 and S2 signals at several energies, in a so-called "Doke plot", a schematic of which is shown in Figure~\ref{fig:doke}, left. If we plot $S2/E_0$ (charge yield, Q$_y$) as a function of $S1/E_0$ (light yield, L$_y$) , we can obtain g$_2$/g$_1$ from the slope, and g$_2$ from the intercept of the fitted line.

Another method to measure g$_1$ and g$_2$ by varying the signal ratio between S1 and S2 is to use different electric fields with one or several calibration sources, exemplified in Figure~\ref{fig:doke}, right. The quantities g$_1$ and g$_2$ are thus detector dependant, for they include the detector's efficiency to detect the light and charge signals.  Finally, the S2 gain can be determined directly from the mean size of the S2 signal for single electrons extracted into the gas phase.  Relative energy resolutions ($\sigma/\mu$) at the level of 4-6\% at energies of a few tens of keV and 2-3\% at energies of a few 100\,keV were reached in xenon TPCs~\cite{Aprile:2011dd,Akerib:2016qlr,XENON:2019dti}.  At higher energies, relevant  for the neutrinoless double beta decay of $^{136}$Xe, relative resolutions of 0.67\%, 0.8\% and 1.2\% around 2.5\,MeV were obtained by LZ~\cite{Pereira:2023rte}, XENON1T~\cite{Aprile:2020yad} and EXO-200~\cite{Anton:2019wmi}, respectively\footnote{Superior energy resolutions can be achieved in high-pressure gas TPCs with electroluminiscent amplification,  see~\cite{Nygren:2009zz}.  As a concrete example, a relative resolution of 0.4\% was obtained in the 5\,kg scale NEXT-White detector~\cite{Renner:2019pfe}, a prototype for NEXT~\cite{Adams:2020cye}.}.

A comprehensive framework to simulate scintillation and ionisation yields and resolutions as a function of interaction type, energy and electric field in a TPC is the Noble Element Simulation Technique (NEST)~\cite{Szydagis:2011tk}. The code, based on phenomenological models informed by a vast array of data, also allows for simulating detector specific effects, once the primary and secondary scintillation gains, as well as the drift field, are specified. NEST models are regularly updated when new data becomes available.

 \section{Past, current and future experiments}
 \label{sec:experiments}
 
The first dual-phase Xe-TPCs that set competitive constraints on WIMP scatters off nuclei were those of the ZEPLIN and XENON programmes, in particular ZEPLIN-II and ZEPLIN-III at the Boulby Mine in the UK, and XENON10 at LNGS in Italy, with first results from ZELPLIN-II and XENON10 in 2007. These initial detectors evolved into LUZ and LUX-ZEPLIN at SURF, USA, and XENON100, XENON1T and XENONnT at LNGS. In parallel, PandaX-I and PandaX-II were constructed at the China Jinping Underground Laboratory (CJPL), with first results in 2014, followed by PandaX-4T. 

\begin{figure}[!h]
\includegraphics[width= 6.8 cm]{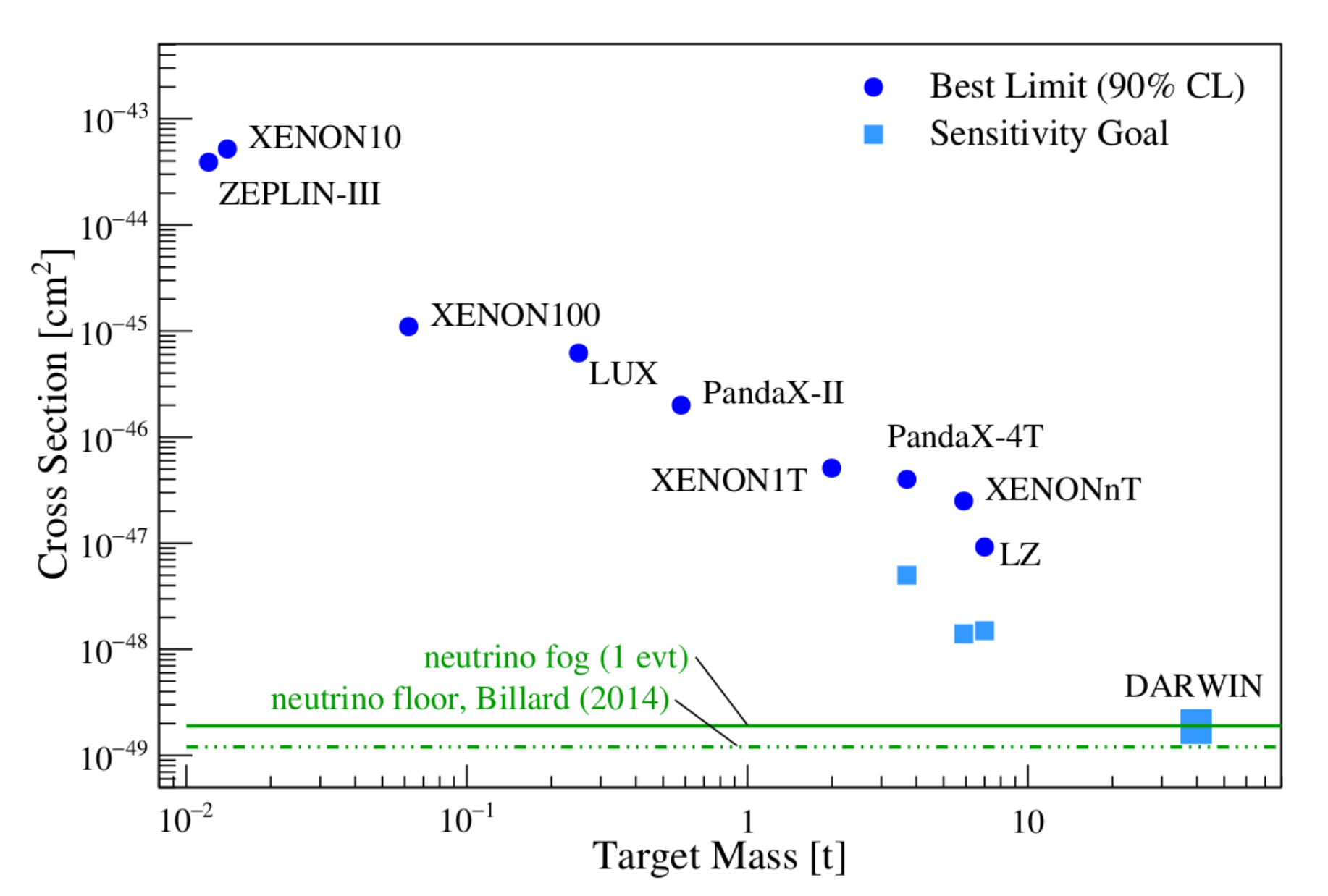}
\includegraphics[width= 6.8 cm]{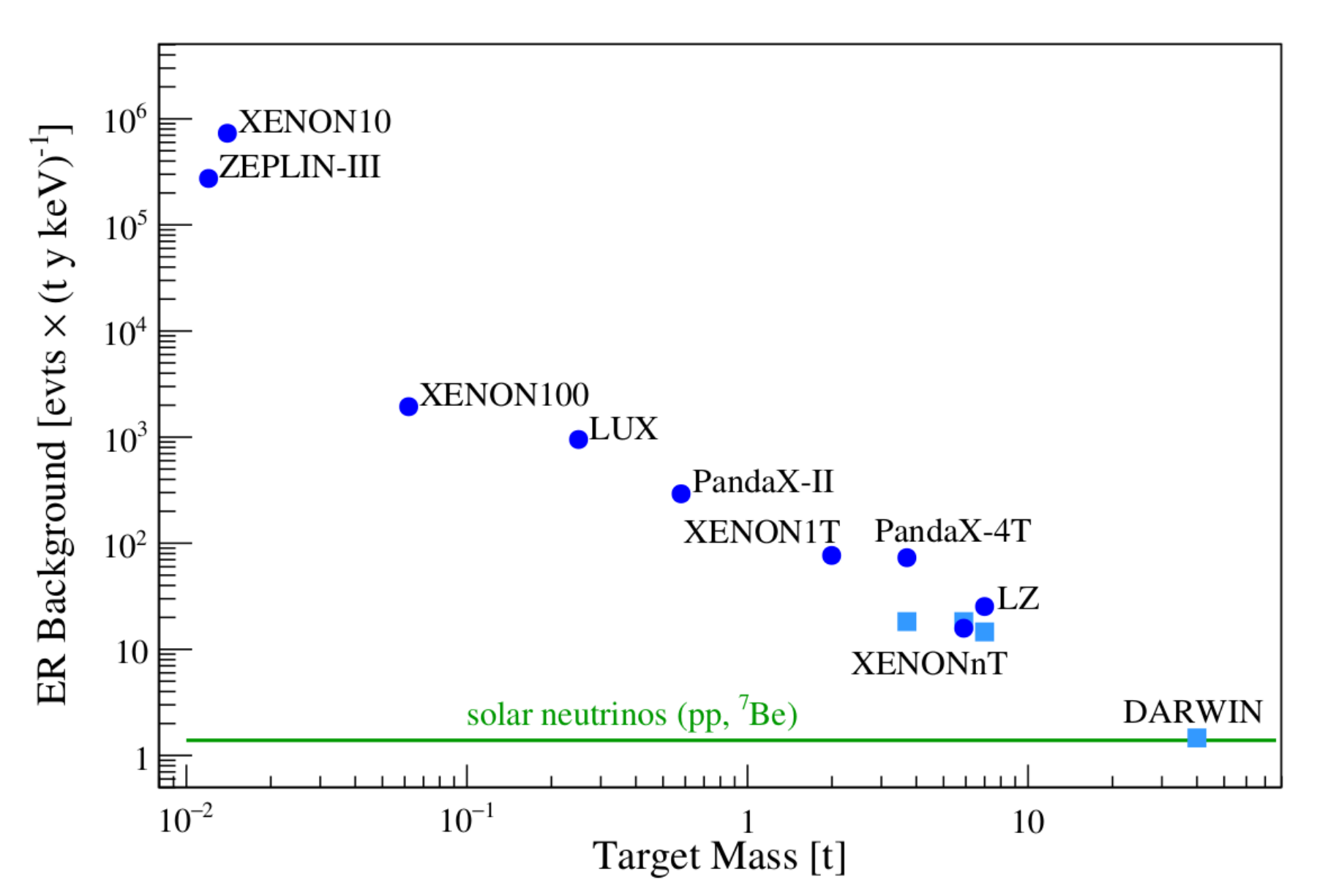}
\caption{Evolution of the sensitivity to SI WIMP-nucleon interactions (left) and electronic recoil background level (right) of dual-phase Xe-TPCs as a function of target mass. We note the logarithmic scales. Figures updated from ~\cite{Aalbers:2022dzr}. }
\label{fig:evolution_time}
\end{figure}

Starting with total masses at the few kilogram and later 100\,kg scale, the detectors evolved and reached target masses at the tonne- and more recently multi-tonne scale. Concomitantly, the background levels in the most inner regions constantly decreased, with now unprecedented electronic recoil levels around 15 events/(t\,y\,keV) in the energy region below 100\,keV.  Figure~\ref{fig:evolution_time} shows the remarkable evolution of the sensitivity to WIMPs (left) and of the background level (right) as a function of target mass. The figures cover a span of about twenty years.
 
 The current generation of detectors employ several tons of LXe: LZ~\cite{LZ:2019sgr}, PandaX-4T~\cite{PandaX-4T:2021bab} and XENONnT~\cite{XENON:2020kmp} have total (target) LXe masses of 10\,t (7\,t), 5.6\, (3.7\,t) and 8.6\,t (5.9\,t), respectively. While their overall TPC design is rather similar, with cylindrical, PTFE enclosed target regions viewed by two arrays of 3-inch diameter Hamamatsu R11410 PMTs, their detailed realisation differs in many technical aspects, some of which are described in the following.
 
\subsection{The LUX-ZEPLIN experiment}
\label{subsec:lz}

The LUX-ZEPLIN (LZ) experiment is located at the Sanford Underground Research Facility (SURF) in South Dakota, USA. The inner, xenon detector, is composed of a TPC and a veto, called Xe Skin, enclosed in a double-walled, low-background titanium cryostat. The dark matter target in the TPC contains 7 tonnes of LXe  in a cylindrical volume which measures 1.5\,m in diameter and height. The vapour phase above the liquid is 8\,mm thick, and the site of the S2 signal production. The TPC is lined with PTFE and instrumented with 494 3-inch Hamamatsu R11410-22 PMTs  arranged in two arrays. The Xe Skin detector, around and underneath the TPC with 2 tonnes of instrumented liquid with 93 1-inch and 38 2-inch PMTs, acts as an anti-coincidence scintillation detector, especially effective for gamma radiation.  The cryostat is surrounded by an outer detector, a system of acrylic tanks containing 17\,t of Gd-loaded liquid scintillator, placed in a 238\,t tank of ultra-pure water. The tank is equipped with 120 8-inch PMTs which record both outer detector and the water Cherenkov signals. LZ has presented first results on WIMP dark matter for an exposure of 60 live days with 5.5\,t of liquid xenon fiducial mass. The main electronic recoil background comes from $^{214}$Pb $\beta$-decays, with a $^{222}$Rn concentration in LXe of 3.3\,$\mu$Bq/kg and from $^{37}$Ar decays, with a half-life of 35\,days. The xenon gas was purified of krypton using gas chromatography, with a $^{\mathrm{nat}}$Kr rate of $(0.14\pm0.02)$\,ppt. 
The  data was consistent with a background-only hypothesis and new upper limits on spin-independent and spin-dependent WIMP-nucleon cross sections for WIMP masses above 9\,GeV were set, with the most stringent constraint at $6.5\times10^{-48}$cm$^2$ for a mass of 30\,GeV/c$^2$~\cite{LZ:2022ufs}. The LZ experiment continues to take science data at SURF, with a projected sensitivity for spin-independent (SI)  WIMP-nucleon cross sections of 1.5$\times$10$^{-48}$\,cm$^2$ for a 40\,GeV/c$^2$ WIMP (at 90\% C.L.) and an exposure of 20\,t\,y.
 
\subsection{The PandaX-4T experiment}
\label{subsec:pandax}
  
The PandaX-4T experiment is located at CJPL, built deep underground in the Jinping Mountains of Sichuan. The inner detector is a TPC with 3.7 tonnes of liquid xenon, contained by 24 PTFE wall panels and observed by 169 and 199 R11410-23 three-inch PMTs in the top and bottom, respectively.  The TPC and xenon are contained in a double-walled, low-background stainless steel cryostat, placed at the centre of a 10\,m diameter and 13\,m tall tank filled with ultra-pure water.  The space between the TPC field cage and the cryostat is instrumented with two rings of one-inch R8520 Hamamatsu PMTs, for an active xenon veto. The xenon inventory was distilled for krypton through a distillation tower prior to and during the commissioning run, which included 95 days of stable data taking, for  a $^{\mathrm{nat}}$Kr rate of $(0.33\pm0.02)$\,ppt.   The dominant electronic recoil background was due to $\beta$-decays from $^3$H, with an average concentration of $5\times10^{-24}$\,mol/mol in Xe, followed by $\beta$-decays from $^{214}$Pb, with a $^{222}$Rn concentration between 4.2 and 5.9\,$\mu$Bq/kg. The commissioning data was used for a WIMP search, with an exposure of 0.63\,tonne\,year. No dark matter candidate events were observed above the expected background, with the most stringent upper limit on SI WIMP-nucleon cross sections of 3.8$\times$10$^{-47}$\,cm$^2$ for a 40\,GeV/c$^2$ WIMP~\cite{PandaX-4T:2021bab}. Due to the tritium contamination of the xenon, likely originating from a CH$_3$T calibration of the previous detector, the collaboration undertook a tritium removal campaign, before the start of the physics data taking. The goal is to reach another order in magnitude improvement in sensitivity with an exposure of 6\,t\,y.
  
\subsection{The XENONnT experiment}
\label{subsec:xenonnt}
 
The XENONnT experiment is located at the INFN Laboratori Nazionali del Gran Sasso (LNGS) in Italy. The TPC contains 5.9\,t of LXe, enclosed by a PTFE cylinder with a diameter of 1.33\,m and a height of 1.49\,m. A total of 494 Hamamatsu R11410-21 3-inch PMTs distributed in a top and a bottom array view the sensitive volume. A double-walled, low-radioactivity stainless steel cryostat houses the TPC and is filled with 8.5\,t of LXe. The cryostat is placed in a 700\,t water Cherenkov muon veto, 10.2\,m high and 9.6\,m in diameter, equipped with 84 8-inch PMTs.  A volume of 33\,m$^3$ of water surrounding the cryostat and delimited by octagonal caps and side reflectors made of high-reflectivity expanded PTFE is used as a neutron veto, seen by 120 8-inch PMTs. In the first physics run, the vetos contained ultra-pure water, while later the water will be doped with Gd to increase the neutron veto efficiency. XENONnT had accumulated 95.1\,days of data during its first science run. A high-throughput radon distillation column was continuously operated to remove radon from the gaseous xenon in the TPC, while a krypton distillation column allowed to achieve a  $^{\mathrm{nat}}$Kr concentration of $<50$\,ppq. In XENONnT, a novel LXe purification system was installed, allowing to reach a very high removal rate of electronegative impurities on a short time scale, with an electron drift lifetime above 10\,ms during the science run.  The dominant electronic recoil background in the first science run  was due to $^{214}$Pb  $\beta$-decays, with a $^{222}$Rn concentration of 1.8\,$\mu$Bq/kg. A blind analysis revealed no significant excess from WIMPs in an exposure of 1.09\,t\,y, with the lowest upper limit on SI WIMP-nucleon cross sections of 2.58$\times$10$^{-47}$\,cm$^2$ at 28\,GeV/c$^2$ (90\% CL)~\cite{XENON:2023sxq}. The ER background of $(15.8\pm1.3)$\,events/(t\,y\,keV) was the lowest in any dark matter detector, and in particular 5 times lower than in XENON1T. XENONnT continues to take data at LNGS, with a further reduced $^{222}$Rn concentration of 0.8\,$\mu$Bq/kg, using the radon distillation system with combined gaseous and liquid xenon flow.

\subsection{Next-generation projects} 
\label{subsec:darwin}

DARWIN, first proposed around 2011~\cite{Baudis:2012bc},  is a next-generation observatory with 40\,t of LXe in the TPC (50\,t total). In the baseline design, the cylindrical TPC, with 2.6\,m diameter and 2.6\,m height, is placed in a low-background, double-walled titanium cryostat, surrounded by active neutron and muon vetos. Two photosensor arrays with a total of 1910 3-inch PMTs will be located at the top and bottom of the TPC, which is lined with high-reflectivity PTFE, surrounded by 92 copper field shaping rings~\cite{Aalbers:2016jon}. The R\&D for new type of photosensors, which could potentially replace the 3-inch PMTs, is ongoing. The studied sensors includes VUV-sensitive silicon photomultipliers, 2-inch$\times$2-inch flat panel PMTs (R12699), and hybrid photosensors.  Recently, the LZ, XENON and DARWIN collaboration joined forces to form the XLZD consortium~\cite{xlzd-website}, with the goal of constructing and operating the next-generation observatory together. The size and scope of the detector might be enlarged, compared to DARWIN, with a 3\,m$\times$3\,m TPC containing 60\,t of LXe (75\,t in total).  The larger mass would allow for a 3-$\sigma$ WIMP discovery at a SI  cross section of 3$\times$10$^{-49}$\,cm$^2$ at 40\,GeV/c$^2$ mass (90\% CL), and would also increase the sensitivity to the neutrinoless double-beta decay of  $^{136}$Xe. The science potential of a large, dual-phase xenon detector is detailed in~\cite{Aalbers:2022dzr}. 

PandaX-xT is the next step in the PandaX programme at CJPL, with $>$30\,t of LXe target in the TPC. Two arrays of Hamamatsu R12699 2-inch PMTs will view the Xe volume. Compared to the 3-inch tubes employed in current TPCs, these new sensors have the advantage of lower radioactivity, faster time response and the possibility of multi-anode readout, with four independent channels per unit. The inner  cryostat vessel will be made of ultra-pure copper, and the space between the inner and outer vessel will contain an active veto. PandaX-xT aims for a 200\,t\,y exposure for WIMPs, and, similar to DARWIN/XLZD, for a broad science reach~\cite{Wang:2023wrr}.

\subsection{Technological challenges} 
\label{subsec:challenges}

\begin{figure}[!h]
\includegraphics[width= 7.1 cm]{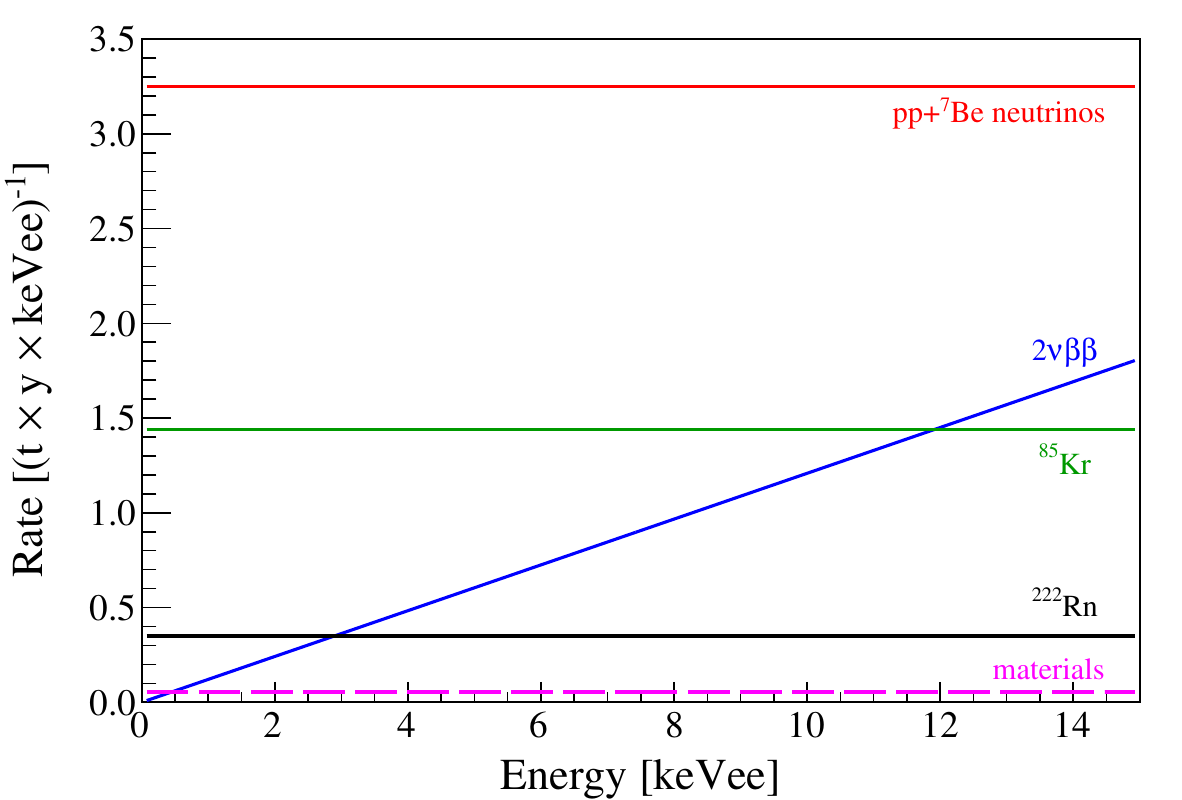}
\includegraphics[width= 6.4 cm]{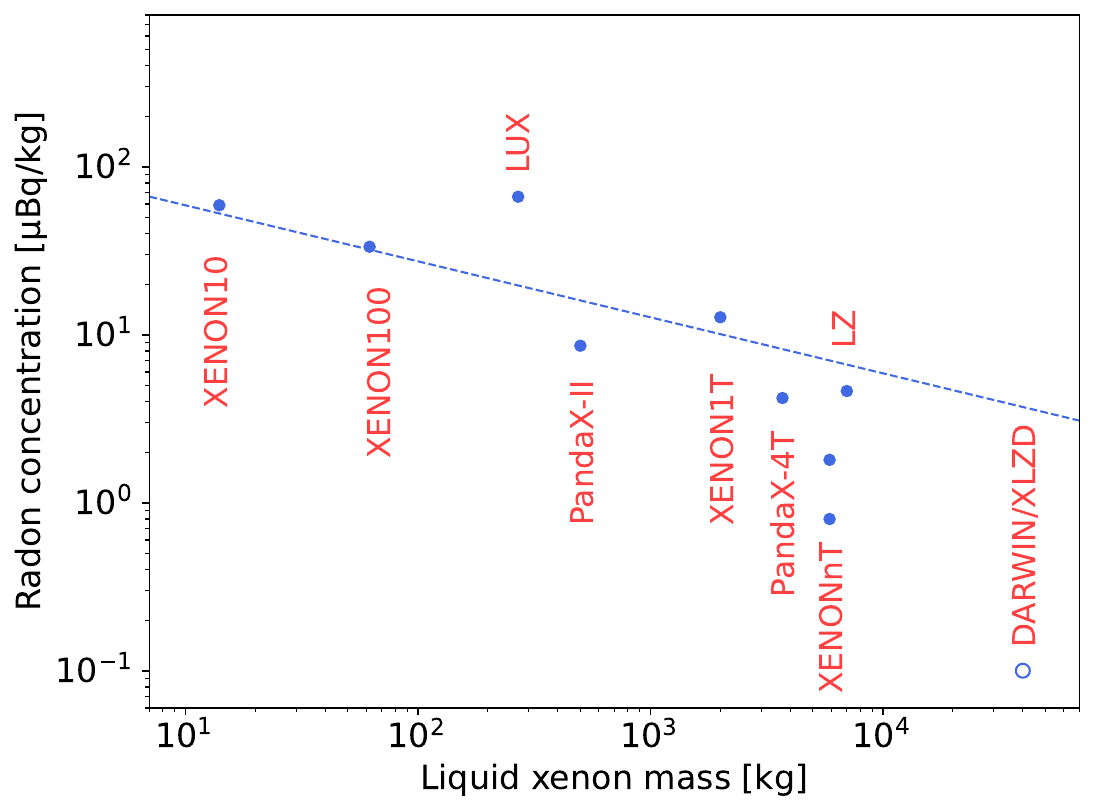}
\caption{(Left): Differential energy spectra of electronic recoil background sources in DARWIN. The dominating contribution is from pp- and $^{7}$Be solar neutrinos. Figure from~\cite{Schumann:2015cpa}. (Right): The evolution of $^{222}$Rn concentration in two-phase Xe-TPCs (measured values, blue dots), together with the expected decrease from the surface-to-volume ratio (dashed line, $x^{-1/3}$). The goal for next-generation TPCs (open circle) is also shown.}
\label{fig:bg_radon_time}
\end{figure}

As we have seen, dual-phase Xe-TPCs were successful in gradually scaling up their target mass from a few kg to multi-tons, and concomitantly reducing the background level for each iteration, while maintaining a low-energy threshold of $\sim$1\,keV for electronic recoils. Notwithstanding, the construction of next-generation detectors with multi-ten-tons target masses poses multiple technological challenges. The background goals are such that electronic and nuclear recoils rates are below the ones from irreducible astrophysical neutrino interactions. This requirement then also sets the goals for the intrinsic $^{222}$Rn and $^{85}$Kr concentrations: the background from the decay of these isotopes should be significantly lower than the solar pp-neutrino elastic scattering rate, as shown in Figure~\ref{fig:bg_radon_time}, left. This translates into 0.1$\mu$Bq/kg for $^{222}$Rn\footnote{A $^{222}$Rn concentration of  0.1$\mu$Bq/kg corresponds to less than one radon atom per 100 mol of xenon. The main background is due to $^{214}$Bi $\beta$-decays which are not accompanied by an $\alpha$-decay and thus cannot be tagged in the TPC.} and 0.1\,ppt for $^{\mathrm{nat}}$Kr, assuming a $^{85}$Kr/$^{\mathrm{nat}}$Kr ratio of 2$\times$10$^{-11}$mol/mol.  $^{\mathrm{nat}}$Kr concentrations of $<50$\,ppq were already achieved by cryogenic distillation~\cite{XENON:2021fkt}, while for $^{222}$Rn a factor of about 10 reduction compared to the current value of 0.8\,$\mu$Bq/kg in XENONnT is still needed, see Figure~\ref{fig:bg_radon_time}, right. This calls for a high liquid xenon throughput (close to 1\,t/hour) with efficient cooling power based on cryogenic heat pumps and radon-free heat exchangers. Cryogenic distillation alone is not sufficient: it must go hand in hand with selection of low radon emanation materials and new coating techniques to prevent radon emanation from surfaces.

Other challenges related to the liquid target are the continuous purification for electronegative impurities and water, to maintain high charge and light yields, as well as new solutions for reliable xenon storage and recuperation at large scales.  Liquid phase purification powered by a liquid xenon pump, as demonstrated in~\cite{Plante:2022khm}, was employed to achieve an electron drift lifetime above 10\,ms in about 8.6\,t of xenon in XENONnT.  A system capable of handling 30\,t of xenon in liquid phase, including long-term storage and and transfer of the cryogenic liquid between storage module and detectors was constructed for PandaX-xT~\cite{Wang:2023wrr}.

Regarding the detector design, electrodes with large ($>2.5$\,m) diameters, with high transparency, minimal sagging and low spurious electron emission, as well as high-voltage feed-throughs that can safely deliver 50\,kV or more to the cathode must be  developed. The LZ collaboration successfully  built custom-woven wire-mesh grids with 1.5\,m diameter, produced  with an in-house built loom to weave the wire meshes~\cite{Stifter:2020ktw}. Finally the cryostat design must be such as to ensure stability, while reducing as much as possible the amount of material, and thus gamma and neutron emitters in proximity to the TPC.

To address a series of challenges related to constructing and operating next-generation Xe-TPCs, several large-scale demonstrators have been built, in particular also Xenoscope, which includes a 2.6\,m tall TPC~\cite{Baudis:2021ipf,Baudis:2023ywo} and Pancake, to deploy a 2.6\, diameter TPC~\cite{xesat2023-website}.

\section{Sensitivity to dark matter candidates}
\label{sec:dm}
 
\begin{figure}[!h]
\centering
\includegraphics[width= 10 cm]{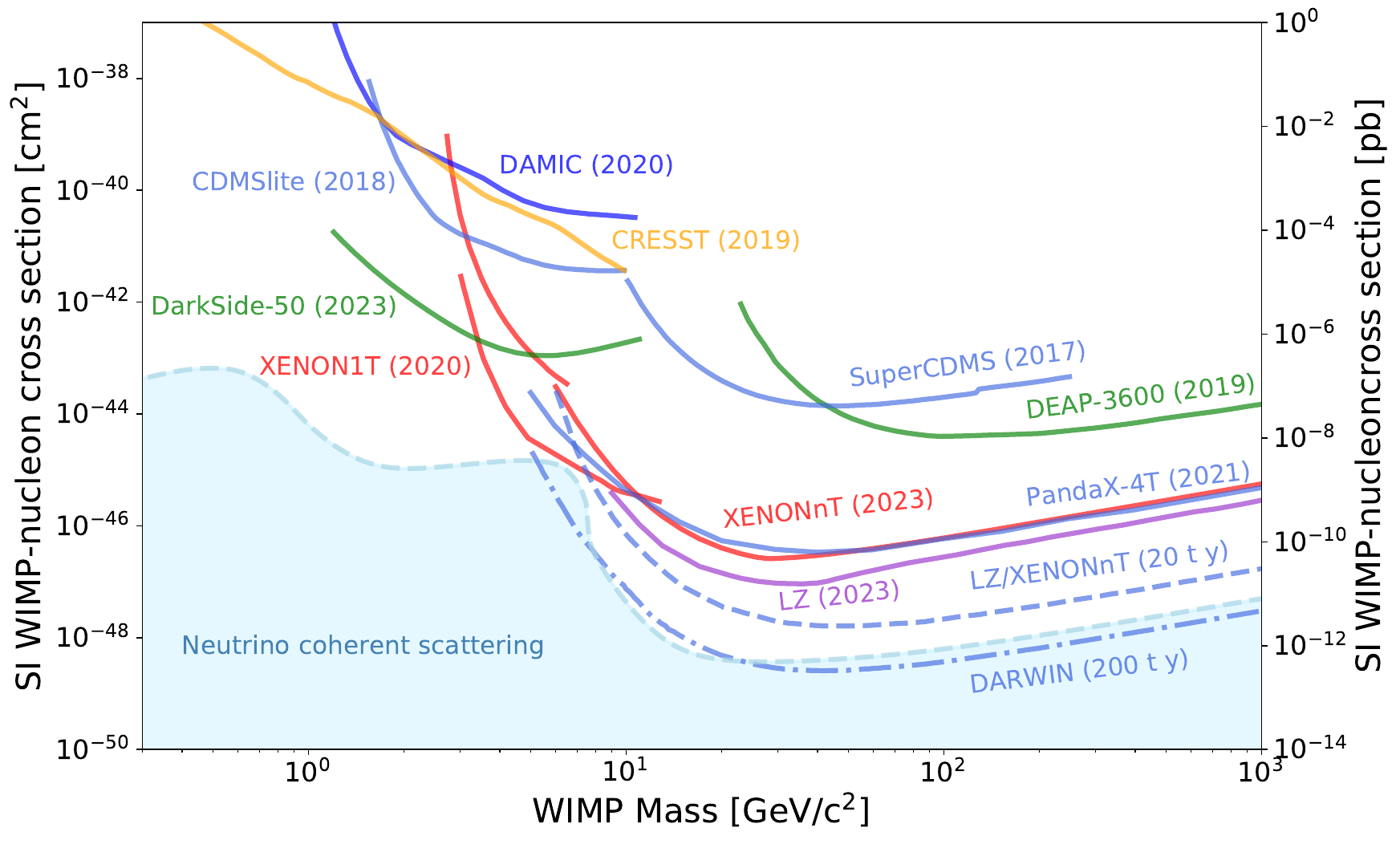}
\caption{Exclusion limits (solid) on the spin-independent WIMP-nucleon cross section from dual-phase Xe-TPCs as well as from other technologies. Projections for LZ and XENONnT (dashed) and for DARWIN/XLZD (dashed-dotted) are also shown. The region where a distinction between a dark matter signal and astrophysical neutrinos will be challenging, albeit not impossible~\cite{OHare:2021utq}, is shown in blue. Figure updated from~\cite{ParticleDataGroup:2022pth}.}
\label{fig:si_limits}
\end{figure}

The current generation of detectors  did not find evidence for WIMP dark matter, but were able to set the world's most stringent limits on WIMP-nucleon interactions over a wide range of particle masses and thus to exclude a large range of possibilities. These results were from very first data and the experiments are yet to reach their design exposures of 20\,t\,y, see Figure~\ref{fig:si_limits}. The search for WIMP dark matter is thus ongoing, with the goal of reaching a sensitivity for the SI WIMP-nucleon cross section around 1.5$\times$10$^{-48}$\,cm$^2$ at 40-50\,GeV/c$^2$ mass~\cite{LZ:2019sgr,XENON:2020kmp}.  While dark matter particles are thought to be electrically neutral, they could posses tiny electromagnetic couplings to photons, allowed by all the data.  The PandaX collaboration looked for nuclear recoils generated by potential dark matter-photon interactions, setting tight constraints on the magnitudes of millicharge, magnetic dipole, electric dipole and anapole moments, as well as first constraints on the charge radius of dark matter~\cite{PandaX:2023toi}. 

\begin{figure}[!h]
\includegraphics[width= 7 cm]{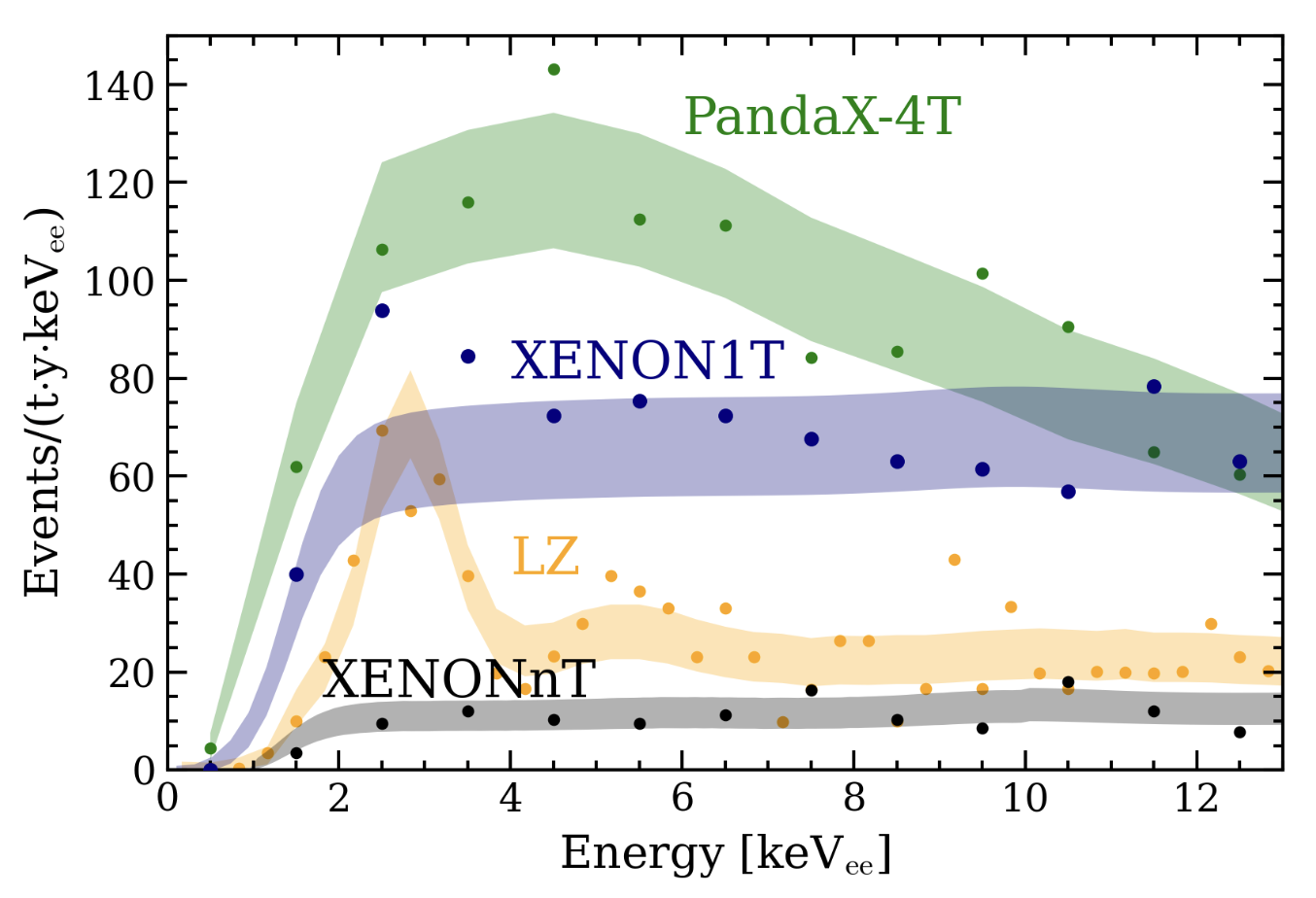}
\includegraphics[width= 6.7 cm]{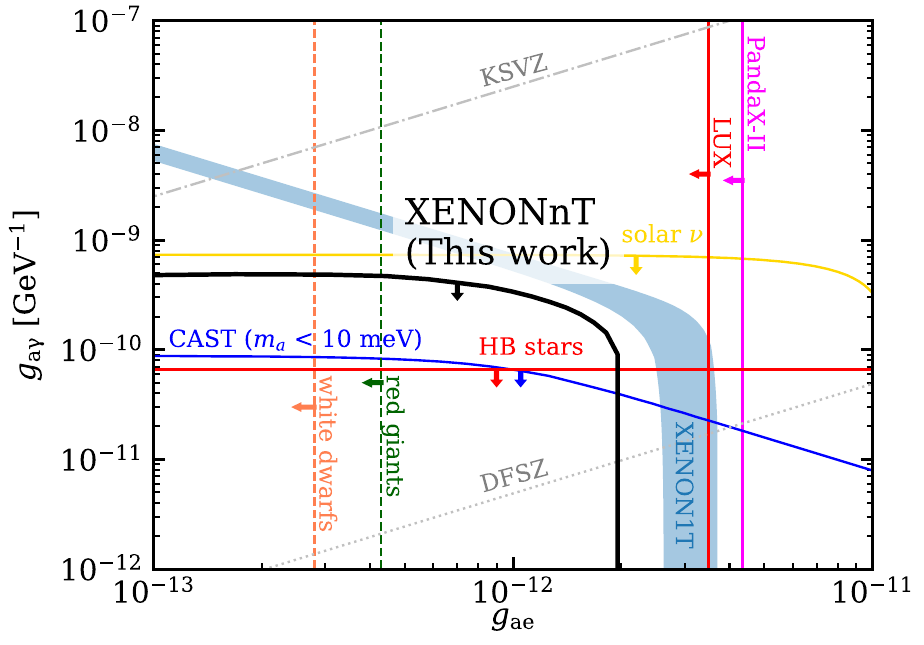}
\caption{(Left): Observed event rates at low energies in XENONnT (grey), compared with XENON1T (blue), as well as with LZ (orange) and PandaX-4T (green). The shaded areas indicate the estimated backgrounds. Figure by J. Ye, Columbia University. (Right): Constraints on solar axion couplings to photons and electrons from XENONnT (black), the XENON1T result (blue), as well as other direct and indirect constraints~\cite{XENONCollaboration:2022kmb}.}
\label{fig:lower_axions}
\end{figure}

Due to unprecedented low electronic-recoil backgrounds and low energy thresholds, high-sensitivity electronic recoil searches also became feasible. These include searches for dark matter electron scattering with particle candidates from a hidden sector, searches for keV-scale axion-like-particles (ALPs) and dark photons via absorption in LXe, and searches for solar axions.  In 2020, XENON1T reported a surprising excess of events in the energy region 1-7\,keV, which could have been due to solar axions (with a statistical significance of  3.4\,$\sigma$), galactic ALPs, or a neutrino magnetic moment. The excess, however, could also have been caused by minute amounts of tritium in the xenon, at the level of 6$\times$10$^{-25}$mol/mol, or less than three tritium atoms per kilogram of LXe~\cite{XENON:2020rca}. With the accumulated number of events, it was not possible to confirm or reject the tritium hypothesis with XENON1T, and a larger detector with lower backgrounds was urgently needed. With  data from its first science run, and an electronic recoil rate five times below the one of its predecessor, XENONnT did not see any excess above background. The rate and shape agreed well with  predictions and, between (1-10)\,keV, was dominated by $^{214}$Pb $\beta$-decays from $^{222}$Rn dissolved in the LXe, followed by interactions of pp solar neutrinos~\cite{XENONCollaboration:2022kmb}\footnote{Most likely the XENON1T low-energy excess  was indeed caused by tritium decays, since in XENONnT the fluid handling and purification is different and measures  to reduce any initial tritium concentration were taken.}.  A comparison between low-energy event rates in XENON1T, showing the excess, XENONnT, PandaX-4T and LZ is shown in  Figure~\ref{fig:lower_axions}, left.  Among other results, XENONnT set new constraints on the coupling of solar axions to electrons and photons, via the axio-electric and inverse Primakoff effects, respectively, shown in Figure~\ref{fig:lower_axions}, right~\cite{XENONCollaboration:2022kmb}. Sensitivity projections for LZ to new physics via low-energy electronic recoils, including mirror and leptophilic dark matter, are detailed in Ref.~\cite{LZ:2021xov}.

\begin{figure}[!h]
\includegraphics[width= 13.6 cm]{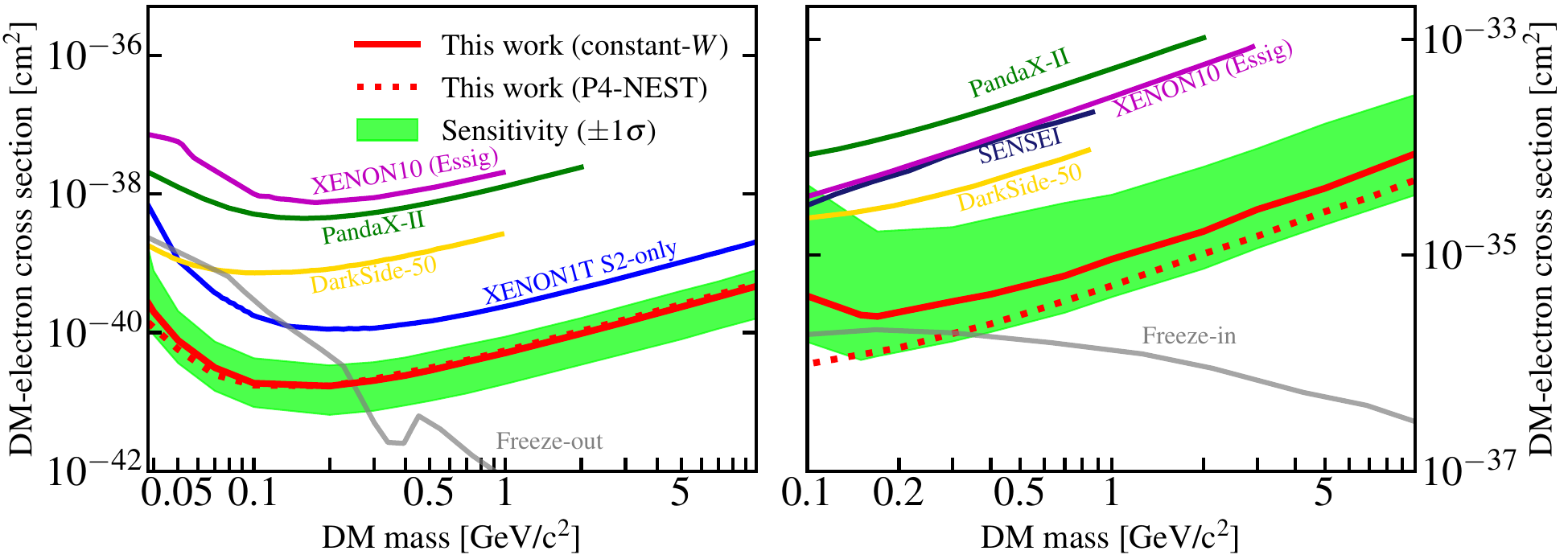}
\caption{Upper limits on the cross section for dark matter electron scatters, assuming a heavy (left, with a dark matter form factor = 1) and light (right, with a dark matter form factor $\sim 1/q^2$) mediator, from the PandaX-4T experiment (red), as well as other constraints. Dual-phase Xe-TPCs thus start testing theoretical predictions, shown as grey bands for different dark matter production scenarios. Figures from~\cite{PandaX:2022xqx}.}
\label{fig:dm_electron_pandax}
\end{figure}

Looking at ionisation-only signals allows for a further reduction of the energy threshold, given the much higher efficiency to detect an ionisation electron compared to a primary scintillation photon (typically 90\% versus 10\%). While this gain comes at the expense of higher backgrounds, it allows nonetheless to set the most stringent limits on light dark matter electron interactions at masses from a few tens of MeV to a few GeV. This is illustrated in Figure~\ref{fig:dm_electron_pandax}, which shows recent result from PandaX-4T~\cite{PandaX:2022xqx} as well as from previous searches.  Dual-phase Xe-TPCs thus start probing theoretical predictions for different dark matter production scenarios in hidden sectors.

 \section{Second order weak interactions}
 \label{sec:doubleweak}
 
Large TPCs using liquid xenon in its natural isotopic abundance are sensitive to double weak decays, such as the double beta decay of $^{134}$Xe and $^{136}$Xe, with Q-values of 825.8\,keV and 2457.8\,keV, respectively, and the double electron capture process in $^{124}$Xe and $^{126}$Xe, at Q-values of 2857\,keV and 920\,keV. In particular, the observation of the neutrinoless decay modes:
\begin{equation}
(Z,A) \rightarrow (Z+2, A) + 2 e^-,
\label{eq:dbd}
\end{equation} 
\begin{equation}
(Z,A) + 2 e^- \rightarrow (Z-2, A),
\label{eq:dec}
\end{equation}
 would be evidence for lepton number violation involving Majorana neutrinos. The observed standard model processes with two neutrinos (so far, $2\nu\beta\beta$ in $^{136}$Xe and $2\nu$ECEC in $^{124}$Xe) allow for comparison to theoretical half-life predictions from nuclear structure calculations in different  nuclear physics models, as well as for new physics searches (see, e.g., \cite{Bossio:2023wpj} for a recent overview).

In its electronic recoil channel, the XENON1T experiment observed for the first time the 2$\nu$ECEC process in  $^{124}$Xe by detecting the simultaneously emitted K-shell X-rays/Auger electrons of the daughter atom $^{124}$Te with a combined energy of 64.33\,keV, which is twice the K-shell binding energy. With a half-life of $(1.1\pm0.2_{\mathrm{stat}}\pm0.1_{\mathrm{sys}})\times$10$^{22}$\,y, this is the slowest process ever measured directly~\cite{XENON:2019dti, XENON:2022evz}.   LZ, PandaX-4T and XENONnT will improve upon these results and will also search for the $2\nu\beta^+$EC decay. This channel has a distinct signature due to the two 511\,keV gammas emitted after the positron annihilates with an electron in the medium. With a predicted half-life around $1.6\times10^{23}$\,y~\cite{Wittweg:2020fak} its first observation is within reach of these running experiments, which will also measure the half-life and in particular also the shape of the $2\nu\beta\beta$-decay of $^{136}$Xe with high statistics and at low energies not previously accessed.

\begin{figure}[!h]
\includegraphics[width= 6.8 cm]{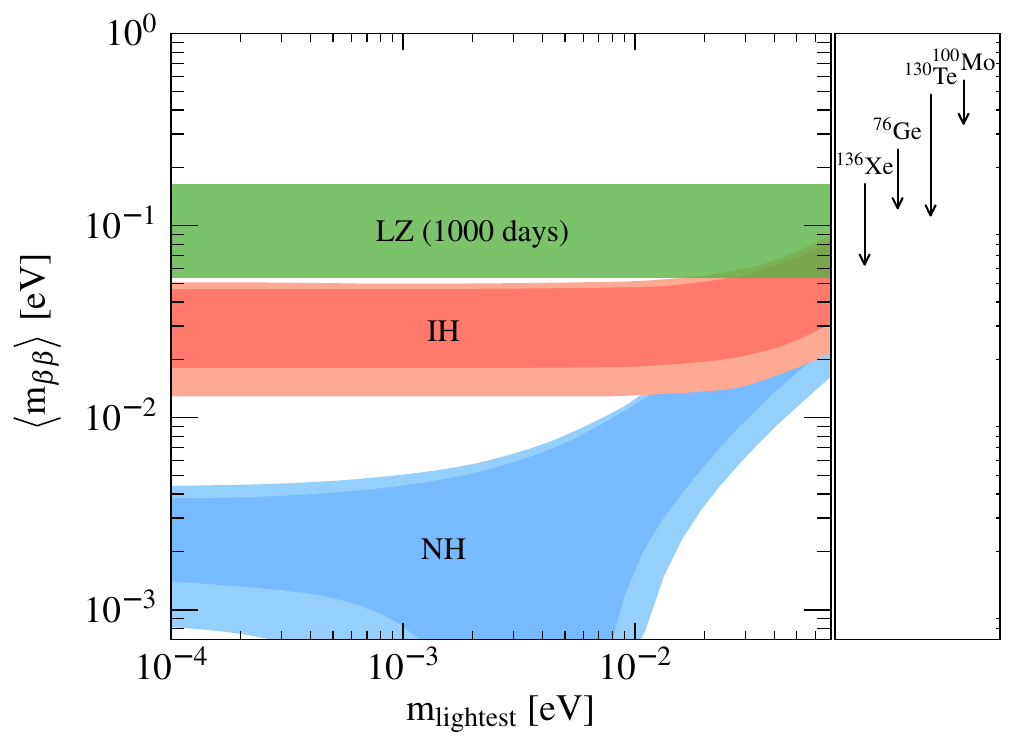}
\includegraphics[width= 6.8 cm]{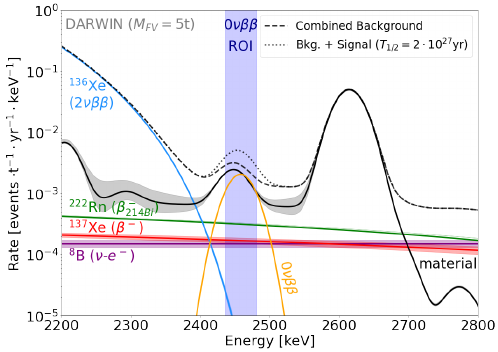}
\caption{(Left): Projected sensitivity of LZ (green band) to the effective Majorana neutrino mass, as a function of the lightest neutrino mass eigenstate. The red and blue contours show the allowed parameter space at 1-sigma for the inverted and normal neutrino mass ordering scenarios, respectively. Figure from~\cite{LZ:2019qdm}. (Right): Predicted background spectrum in DARWIN  around the region-of-interest for $0\nu\beta\beta$-decay for 5\,t LXe fiducial mass. A hypothetical signal of 0.5\,events/y, corresponding to a half-life of $2\times 10^{27}$\,y, is shown for comparison.  Figure updated from~\cite{Agostini:2020adk}.}
\label{fig:doublebeta}
\end{figure}

Finally, the current generation of detectors will set constraints on the $0\nu\beta\beta$-decays of $^{136}$Xe and $^{134}$Xe, as predicted or shown in \cite{PandaX:2022kwg,XENON:2022evz, LZ:2019qdm, LZ:2021blo}, see Figure~\ref{fig:doublebeta}, left. In spite of relative energy resolutions ($\sigma/\mu$) of 0.8\% and 0.67\%~\cite{Aprile:2020yad,Pereira:2023rte} at high energies,  the attainable half-lives are not competitive to dedicated experiments such as KamLAND-Zen~\cite{KamLAND-Zen:2022tow} and the proposed nEXO~\cite{nEXO:2021ujk}. Notwithstanding, the analyses will deliver proof-of-principle methods towards higher sensitivity searches in the DARWIN/XLZD and PandaX-xT detectors.  As an example DARWIN, with 40\,t of natural xenon in the TPC, is predicted to achieve a sensitivity of $3\times10^{27}$\,y (90\% C.L.) after ten years of operation~\cite{Agostini:2020adk}.  The predicted background spectrum around the Q-value of the decay, together with a hypothetical signal, is shown in  see Figure~\ref{fig:doublebeta}, right. With an enlarged xenon mass of 60\,t in the TPC, as advocated by XLZD, a half-life sensitivity of $\sim$$5\times10^{27}$\,y (90\% C.L.)  could be reached after ten years, allowing to fully probe the inverted mass ordering scenario for neutrinos.

\section{Neutrino detection}
 \label{sec:neutrinos}

Recently, dual-phase Xe-TPCs have reached target scales which allow them to explore a variety of astrophysical neutrino sources. The main detection channels are elastic neutrino-electron scattering and coherent elastic neutrino-nucleus scattering:
\begin{equation}
\nu_x + e^- \rightarrow \nu_x + e^-,
\label{eq:nue}
\end{equation} 
\begin{equation}
\nu_x + (Z,A) \rightarrow  \nu_x + (Z, A),
\label{eq:cevens}
\end{equation}
where $\nu_x$ stands for all three neutrino flavours, $\nu_{e}, \nu_{\mu}, \nu_{\tau}$. The latter process, in which the neutrino scatters off a nucleus via Z-boson exchange and the nucleus recoils a a whole, is coherent up to neutrino energies around 50\,MeV.

The observation of elastic neutrino-electron scattering of solar $pp$ and $^{7}$Be neutrinos will allow to measure the $pp$ and  $^{7}$Be fluxes with $<$1\% statistical precision for a 300\,t\,y exposure, with rates of 365 events/(t\,y) and 140 events/(t\,y), respectively. This will imply a first measurement of the electron neutrino survival probability with 4\% relative uncertainty, as well as of the weak mixing angle, with  5\% relative uncertainty, at energies below 200\,keV~\cite{DARWIN:2020bnc}. The main backgrounds for these measurements are $^{222}$Rn decays, with a required concentration of 0.1\,$\mu$Bq/kg, and the $2\nu\beta\beta$ decay of $^{136}$Xe. A first detection of solar $pp$ neutrinos will already be achievable with the current generation of detectors. As we have seen, the solar $pp$ rate was almost half the rate from $^{214}$Bi $\beta$-decays in XENONnT's first science run, in the energy range $(1-10)$\,keV. It is also interesting to note that, for the first time in a dark matter detector, the neutrino-induced rate is a factor of $\sim$1.6 above the one from the radioactivity of detector materials~\cite{XENONCollaboration:2022kmb}.

Coherent elastic neutrino-nucleus scattering (CE$\nu$NS) was proposed fifty years ago by Daniel Z. Freedman~\cite{Freedman:1973yd}, who wrote: "The experiments are very difficult, although the estimated cross sections (about 10$^{-38}$\,cm$^2$ on carbon) are favorable."\footnote{The manuscript was received in October 1973, only a few months after the Gargamelle collaboration had presented the first evidence of the weak neutral current in July of the same year~\cite{GargamelleNeutrino:1973jyy}.} The experiments were difficult indeed, and the process was only observed four decades later by the COHERENT collaboration with a CsI(Na) scintillating crystal~\cite{COHERENT:2017ipa}, and later also with a liquid argon detector~\cite{COHERENT:2020iec}. COHERENT uses neutrinos from pion decays at rest, produced at the Spallation Neutron Source at Oak Ridge National Laboratory. Thus, to this day, CE$\nu$NS was never observed on xenon and never with "naturally occurring" neutrinos.  

\begin{figure}[!h]
\includegraphics[width= 5.7 cm]{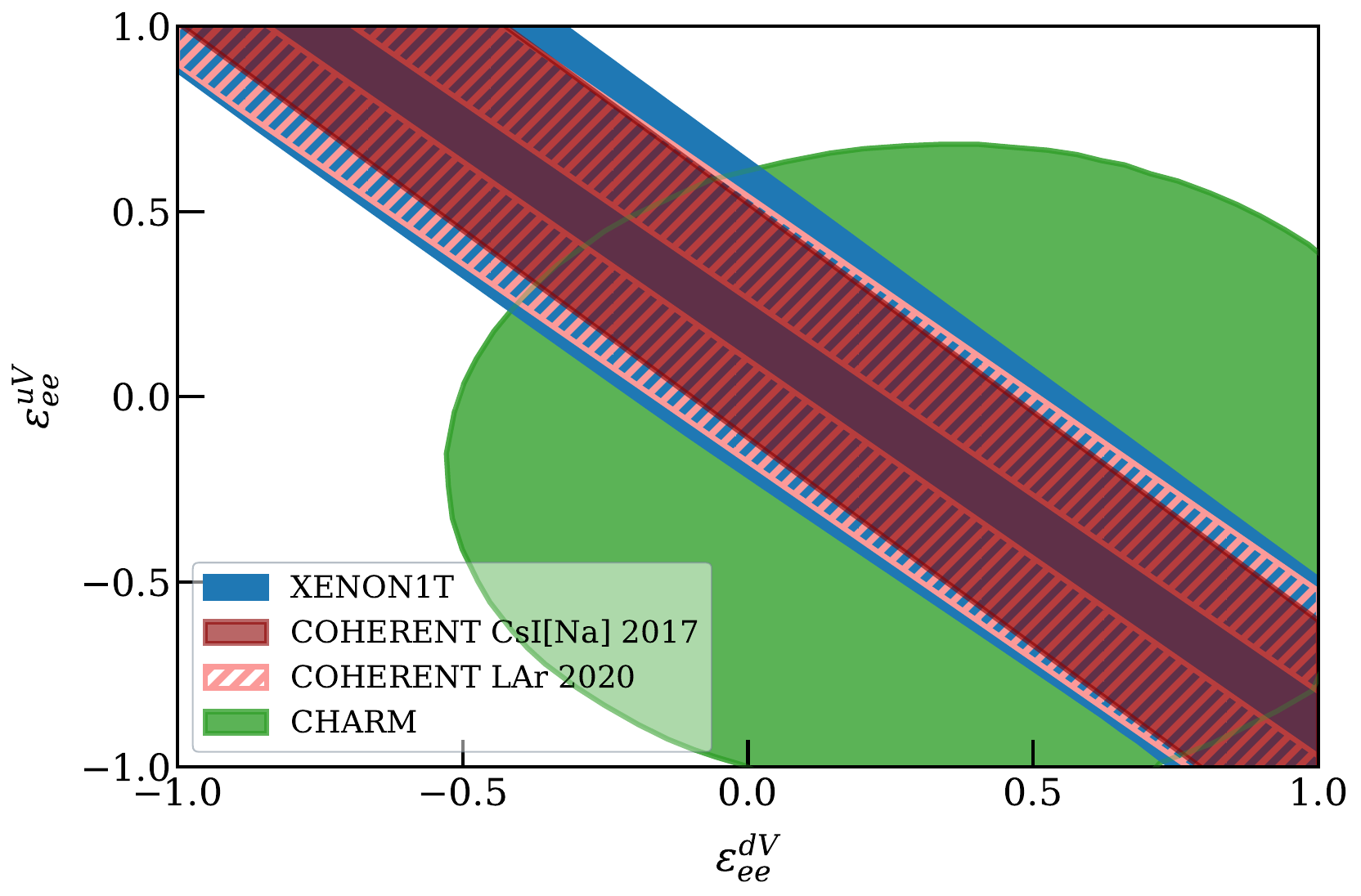}
\includegraphics[width= 7.8 cm]{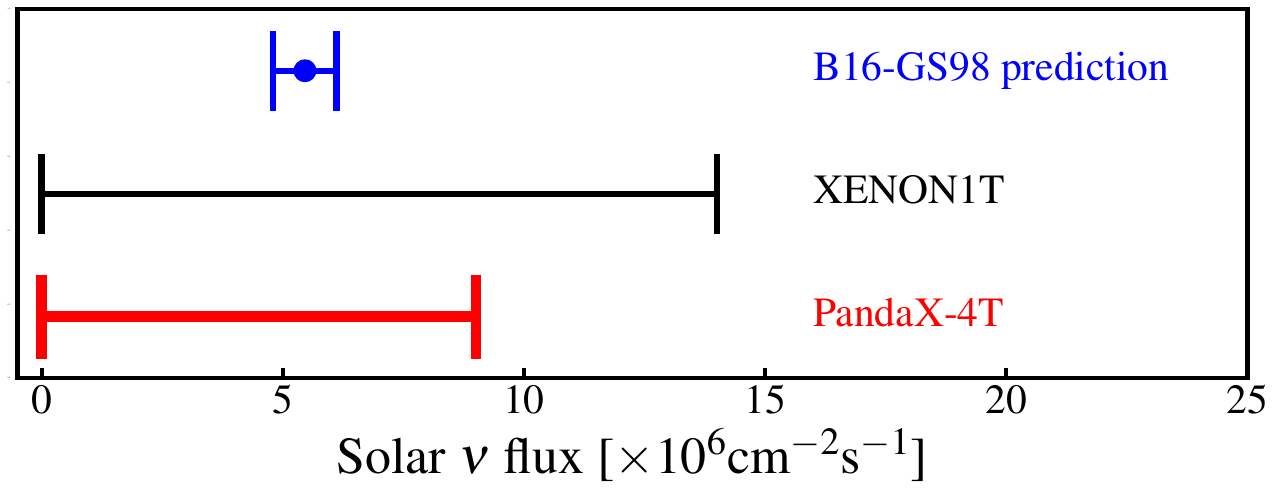}
\caption{(Left): Constraints on non-standard vector couplings between electron neutrinos and up/down quarks from XENON1T (blue), COHERENT (light and dark red) and CHARM (green). Shown is the ratio with respect to the SM expectation: the presence of NSI would result in an enhancement or a suppression of the CE$\nu$NS rate. Figure from~\cite{XENON:2020gfr}. (Right): Constraints on the $^8$B solar neutrino flux via CE$\nu$NS from XENON1T (black), PandaX-4T (red) as well as the standard solar model (B16-GS98) prediction (blue). Figure from~\cite{PandaX:2022aac}.}
\label{fig:boron8}
\end{figure}

The cross section is proportional  to the number of neutrons squared, hence heavy nuclei  are ideal to observe astrophysical neutrinos via this reaction. For $^8$B solar neutrinos,  about 99\% of events are at nuclear recoil energies below 3\,keV in a xenon detector, which poses one of the main difficulties in observing CE$\nu$NS-induced events.  Firstly, the detection efficiencies in Xe-TPCs are relatively low at these energies (in particular if both S1 and S2 signals are required) and secondly, the light and charge yields have large associated uncertainties\footnote{We refer to Ref.~\cite{Xiang:2023csc} for a recent study  of uncertainties of low-energy LXe responses to nuclear recoils and their impact on solar $^8$B neutrino searches.}. In addition, the combinatorial background (random pairing of S1 and S2 events) is largest at low energies~\cite{XENON:2020gfr}, and the problem is more severe for larger detectors with increased electron drift time (and possibly higher rates of spurious events). Nonetheless, the current generation of multi-ton xenon TPCs aim to observe  $^8$B solar neutrinos for the first time. Figure~\ref{fig:boron8} show constraints on the $^8$B  neutrino flux from XENON1T~\cite{XENON:2020gfr} and PandaX-4T~\cite{PandaX:2022aac}, as well as constraints on non-standard neutrino interactions from XENON1T and COHERENT. Next-generation experiments  will carry out precision measurements of $^8$B neutrinos induced CE$\nu$NS, and, with an event rate of about 90 events per tonne and year~\cite{Baudis:2013qla} they will  probe non-standard neutrino interactions and provide an independent measurement of the solar  $^8$B neutrino flux~\cite{Aalbers:2022dzr}.

\begin{figure}[!h]
\includegraphics[width= 6.5 cm]{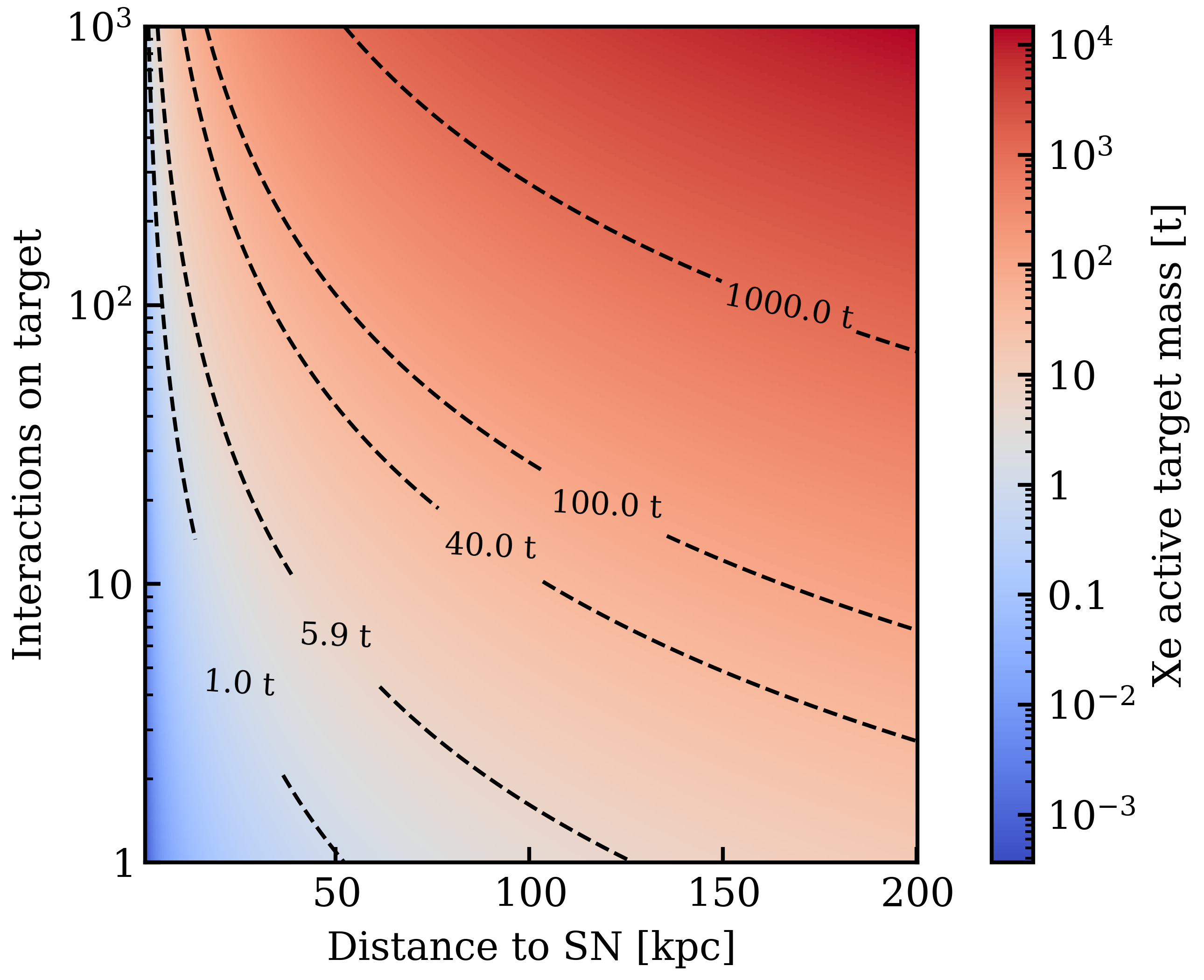}
\includegraphics[width= 6.0 cm]{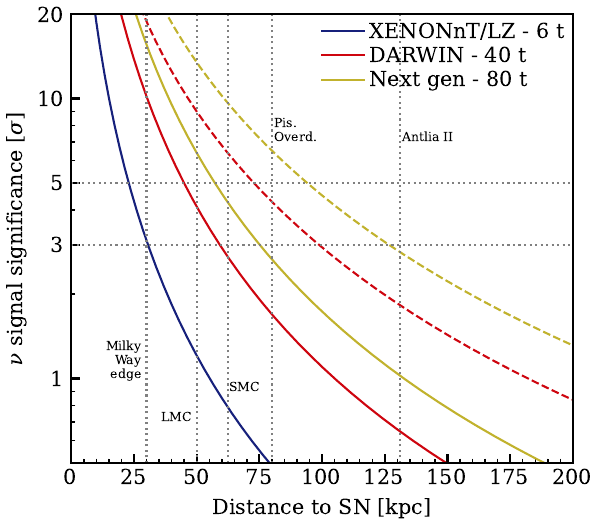}
\caption{Number of SN $\nu$ interactions in a Xe detector for a range of target masses (left) and detection significance (right) as a function of distance to the SN, for a 27\,M$_\odot$ progenitor. The solid (dashed) curves in the right plot, also showing the sensitivity of XENONnT/LZ, assume current (decreased by a factor of 10) level of backgrounds. Figures from~\cite{PeresRicardoThesis2023}.}
\label{fig:sn-neutrinos}
\end{figure}

Apart from solar neutrinos, two-phase Xe-TPCs will be able to detect CE$\nu$NS-induced events from supernova (SN) neutrinos.  During the collapse of a star, about 99\% of the gravitational binding energies of the proto-neutron star goes into neutrinos of all flavours, with energies up to tens of MeV. Given the nature of the interaction process, xenon detectors are sensitive to all neutrino flavours with few events per tonne expected for a core-collapse SN at a distance of 10\,kpc. Assuming a SN with 27\,M$_\odot$ progenitor at a distance of 10\,kpc, 700 events are expected in the DARWIN detector with 40\,t of LXe~\cite{Lang:2016zhv}. This number increases for a larger mass, as does the significance of a SN detection, for a given distance, as illustrated in Figure~\ref{fig:sn-neutrinos}. 

Current xenon experiments participate in the supernova early warning system (SNEWS) network~\cite{snews-website}, which prepares and provides an early warning system for galactic SN to facilitate the observation of optical counterparts. Finally, next-generation Xe-TPCs  will  start constraining the diffuse supernova background (DSNB). Understanding core-collapse SN depends on probing the DSNB with all neutrino flavours, and currently only upper limits on the $\nu_e, \bar{\nu}_e$ flux the SNO and Super-Kamiokande experiments exist, around 19\,cm$^{-2}$s$^{-1}$ and 2.7\,cm$^{-2}$s$^{-1}$, respectively. Limits on the fluxes of $\nu_{\mu,\tau}, \bar{\nu}_{\mu,\tau}$  are much weaker, around 10$^3$\,cm$^{-2}$s$^{-1}$, and a large xenon detector could improve these by about a factor of 100. While this would help in constraining SN models, the actual detection of the DSNB  remains challenging, even with a 1000\,t\,y exposure~\cite{Suliga:2021hek}.

\section{Conclusion}
\label{sec:conclusions}

Dual-phase Xe-TPCs are relatively new detectors in the landscape of astroparticle physics experiments. Developed in the early twenty first century primarily to search for dark matter in the form of WIMPs, they soon surpassed other technologies in terms of their sensitivity to WIMP-nucleon interactions over a large range of WIMP masses~\cite{Baudis:2012ig}. Almost twenty years later, detectors using several tonnes of liquid xenon can observe signals down to a few quanta (photons and electrons) with unprecedented low background rates. These detectors will soon observe for the first time solar neutrinos at energies not previously explored, both via neutrino-electron elastic scattering and via CE$\nu$NS processes. While solar and other astrophysical neutrinos will ultimately limit the sensitivity to dark matter, they bring forth exciting signatures in their own right. Other compelling searches are those for ultra-rare, second order weak processes, for these provide unique probes of beyond standard model physics. While the current generation of detectors (LZ, PandaX-4T and XENONnT) continue to acquire data in different deep underground laboratories, next-generation experiments at the multi-ten-ton scale (DARWIN/XLZD and PandaX-xT) are being planned, with large-scale R\&D projects and prototyping ongoing. With foreseen first data in the early 2030s, these detectors might discover dark matter particles, and thus solve a problem which is almost a century old. At the same time, they will break new grounds in other areas of astroparticle physics, in particular in neutrino physics and rare nuclear decays. Certainly this will not be the final generation of Xe-TPCs. To make sense of future observations, including those which are unexpected, likely a series of new, larger and more sophisticated experiments will need to be designed.  To paraphrase the authors of Ref.~\cite{Aprile:2008bga}, we believe that the best pages in the history of dual-phase Xe-TPCs are yet to be written.

\ack{This work was supported by the University of Zurich, by the European Research Council (ERC) under the European Union's Horizon 2020 research and innovation programme, grant agreement No. 742789 ({\sl Xenoscope}), by the SNF grant 20FL20-201437, as well as by the European Union's Horizon 2020 research and innovation programme under the Marie Sk\l{}odowska-Curie grant agreement No 860881-HIDDeN. We thank Michael Murra for providing the data for Figure~\ref{fig:bg_radon_time}, right, and M. Schumann for the updated Figure~\ref{fig:evolution_time}. }


\bibliographystyle{RS}
\bibliography{baudis_tpc}

\begin{thebibliography}{99}

\bibitem{Nygren:2018sjx}
Nygren DR. 2018  {Origin and development of the TPC idea}. {\em Nucl.\
  Instrum.\ Meth.\ A} \textbf{907}, 22--30.
(\href{http://dx.doi.org/10.1016/j.nima.2018.07.015}{10.1016/j.nima.2018.07.015})

\bibitem{Nygren:1976fe}
Nygren D. 1974  {The Time Projection Chamber: A New 4 pi Detector for Charged
  Particles}. {\em eConf} \textbf{C740805}, 58.

\bibitem{Aprile:2009dv}
Aprile E, Doke T. 2010  {Liquid Xenon Detectors for Particle Physics and
  Astrophysics}. {\em Rev.\ Mod.\ Phys.} \textbf{82}, 2053--2097.
(\href{http://dx.doi.org/10.1103/RevModPhys.82.2053}{10.1103/RevModPhys.82.2053})

\bibitem{Chepel:2012sj}
Chepel V, Araujo H. 2013  {Liquid noble gas detectors for low energy particle
  physics}. {\em JINST} \textbf{8}, R04001.
(\href{http://dx.doi.org/10.1088/1748-0221/8/04/R04001}{10.1088/1748-0221/8/04/R04001})

\bibitem{Gonzalez-Diaz:2017gxo}
Gonzalez-Diaz D, Monrabal F, Murphy S. 2018  {Gaseous and dual-phase time
  projection chambers for imaging rare processes}. {\em Nucl.\ Instrum.\ Meth.\
  A} \textbf{878}, 200--255.
(\href{http://dx.doi.org/10.1016/j.nima.2017.09.024}{10.1016/j.nima.2017.09.024})

\bibitem{Aprile:2008bga}
Aprile E, Bolotnikov AE, Bolozdynya AL, Doke T. 2008 {\em {Noble Gas
  Detectors}}.
Wiley.
(\href{http://dx.doi.org/10.1002/9783527610020}{10.1002/9783527610020})

\bibitem{Bolozdynya:2010zz}
Bolozdynya AI. 2010 {\em {Emission detectors}}.
World Scientific.

\bibitem{Albert:2013gpz}
Albert J et~al.. 2014  {Improved measurement of the $2\nu\beta\beta$ half-life
  of $^{136}$Xe with the EXO-200 detector}. {\em Phys.\ Rev.\ C} \textbf{89},
  015502.
(\href{http://dx.doi.org/10.1103/PhysRevC.89.015502}{10.1103/PhysRevC.89.015502})

\bibitem{KamLAND-Zen:2019imh}
Gando A et~al.. 2019  {Precision measurement of the $^{136}$Xe two-neutrino
  $\beta\beta$ spectrum in KamLAND-Zen and its impact on the quenching of
  nuclear matrix elements}. {\em Phys.\ Rev.\ Lett.} \textbf{122}, 192501.
(\href{http://dx.doi.org/10.1103/PhysRevLett.122.192501}{10.1103/PhysRevLett.122.192501})

\bibitem{XENON:2019dti}
Aprile E et~al.. 2019  {Observation of two-neutrino double electron capture in
  $^{124}$Xe with XENON1T}. {\em Nature} \textbf{568}, 532--535.
(\href{http://dx.doi.org/10.1038/s41586-019-1124-4}{10.1038/s41586-019-1124-4})

\bibitem{XENON:2022evz}
Aprile E et~al.. 2022  {Double-Weak Decays of $^{124}$Xe and $^{136}$Xe in the
  XENON1T and XENONnT Experiments}. {\em Phys. Rev. C} \textbf{106}, 024328.
(\href{http://dx.doi.org/10.1103/PhysRevC.106.024328}{10.1103/PhysRevC.106.024328})

\bibitem{EXO-200:2017vqi}
Albert JB et~al.. 2017  {Searches for double beta decay of $^{134}$Xe with
  EXO-200}. {\em Phys. Rev. D} \textbf{96}, 092001.
(\href{http://dx.doi.org/10.1103/PhysRevD.96.092001}{10.1103/PhysRevD.96.092001})

\bibitem{XMASS:2018txy}
Abe K et~al.. 2018  {Improved search for two-neutrino double electron capture
  on $^{124}$Xe and $^{126}$Xe using particle identification in XMASS-I}. {\em
  PTEP} \textbf{2018}, 053D03.
(\href{http://dx.doi.org/10.1093/ptep/pty053}{10.1093/ptep/pty053})

\bibitem{Platzman:1961}
Platzman R. 1961  {Total ionization in gases by high-energy particles: An
  appraisal of our understanding}. {\em The International Journal of Applied
  Radiation and Isotopes} \textbf{10}, 116 -- 127.
(\href{http://dx.doi.org/https://doi.org/10.1016/0020-708X(61)90108-9}{https://doi.org/10.1016/0020-708X(61)90108-9})

\bibitem{DahlThesis}
Dahl CE. 2009 {\em {The Physics of Background Discrimination in Liquid Xenon,
  and First Results from XENON10 in the Hunt for WIMP Dark Matter}}.
PhD thesis Princeton University.

\bibitem{Baudis:2021dsq}
Baudis L, Sanchez-Lucas P, Thieme K. 2021  {A measurement of the mean
  electronic excitation energy of liquid xenon}. {\em Eur. Phys. J. C}
  \textbf{81}, 1060.
(\href{http://dx.doi.org/10.1140/epjc/s10052-021-09834-x}{10.1140/epjc/s10052-021-09834-x})

\bibitem{Anton:2019hnw}
Anton G et~al.. 2020  {Measurement of the scintillation and ionization response
  of liquid xenon at MeV energies in the EXO-200 experiment}. {\em Phys. Rev.
  C} \textbf{101}, 065501.
(\href{http://dx.doi.org/10.1103/PhysRevC.101.065501}{10.1103/PhysRevC.101.065501})

\bibitem{Hitachi:2005ti}
Hitachi A. 2005  {Properties of liquid xenon scintillation for dark matter
  searches}. {\em Astropart. Phys.} \textbf{24}, 247--256.
(\href{http://dx.doi.org/10.1016/j.astropartphys.2005.07.002}{10.1016/j.astropartphys.2005.07.002})

\bibitem{Lindhard:1963}
J.~Lindhard, V.~Nielsen MS, Thomsen P. 1963  {Integral equations governing
  radiation effects}. {\em Kgl. Danske Videnskab., Selskab. Mat. Fys. Medd}
  \textbf{33}.

\bibitem{Sorensen:2011bd}
Sorensen P, Dahl CE. 2011  {Nuclear recoil energy scale in liquid xenon with
  application to the direct detection of dark matter}. {\em Phys. Rev. D}
  \textbf{83}, 063501.
(\href{http://dx.doi.org/10.1103/PhysRevD.83.063501}{10.1103/PhysRevD.83.063501})

\bibitem{Lenardo:2014cva}
Lenardo B, Kazkaz K, Manalaysay A, Mock J, Szydagis M, Tripathi M. 2015  {A
  Global Analysis of Light and Charge Yields in Liquid Xenon}. {\em IEEE Trans.
  Nucl. Sci.} \textbf{62}, 3387--3396.
(\href{http://dx.doi.org/10.1109/TNS.2015.2481322}{10.1109/TNS.2015.2481322})

\bibitem{Baudis:2023ywo}
Baudis L et~al.. 2023  {Electron transport measurements in liquid xenon with
  Xenoscope, a large-scale DARWIN demonstrator}. .

\bibitem{Aprile:2011dd}
Aprile E et~al.. 2012  {The XENON100 Dark Matter Experiment}. {\em Astropart.
  Phys.} \textbf{35}, 573--590.
(\href{http://dx.doi.org/10.1016/j.astropartphys.2012.01.003}{10.1016/j.astropartphys.2012.01.003})

\bibitem{Akerib:2016qlr}
Akerib DS et~al.. 2017  {Signal yields, energy resolution, and recombination
  fluctuations in liquid xenon}. {\em Phys. Rev. D} \textbf{95}, 012008.
(\href{http://dx.doi.org/10.1103/PhysRevD.95.012008}{10.1103/PhysRevD.95.012008})

\bibitem{Pereira:2023rte}
Pereira G, Silva C, Solovov VN. 2023  {Energy resolution of the LZ detector for
  high-energy electronic recoils}. {\em JINST} \textbf{18}, C04007.
(\href{http://dx.doi.org/10.1088/1748-0221/18/04/C04007}{10.1088/1748-0221/18/04/C04007})

\bibitem{Aprile:2020yad}
Aprile E et~al.. 2020  {Energy resolution and linearity of XENON1T in the MeV
  energy range}. {\em Eur. Phys. J. C} \textbf{80}, 785.
(\href{http://dx.doi.org/10.1140/epjc/s10052-020-8284-0}{10.1140/epjc/s10052-020-8284-0})

\bibitem{Anton:2019wmi}
Anton G et~al.. 2019  {Search for Neutrinoless Double-$\beta$ Decay with the
  Complete EXO-200 Dataset}. {\em Phys. Rev. Lett.} \textbf{123}, 161802.
(\href{http://dx.doi.org/10.1103/PhysRevLett.123.161802}{10.1103/PhysRevLett.123.161802})

\bibitem{Nygren:2009zz}
Nygren D. 2009  {High-pressure xenon gas electroluminescent TPC for 0nu beta
  beta-decay search}. {\em Nucl. Instrum. Meth. A} \textbf{603}, 337--348.
(\href{http://dx.doi.org/10.1016/j.nima.2009.01.222}{10.1016/j.nima.2009.01.222})

\bibitem{Renner:2019pfe}
Renner J et~al.. 2019  {Energy calibration of the NEXT-White detector with 1\%
  resolution near Q$_{\beta \beta}$ of$^{136}$Xe}. {\em JHEP} \textbf{10}, 230.
(\href{http://dx.doi.org/10.1007/JHEP10(2019)230}{10.1007/JHEP10(2019)230})

\bibitem{Adams:2020cye}
Adams C et~al.. 2020  {Sensitivity of a tonne-scale NEXT detector for
  neutrinoless double beta decay searches}. .

\bibitem{Szydagis:2011tk}
Szydagis M, Barry N, Kazkaz K, Mock J, Stolp D, Sweany M, Tripathi M, Uvarov S,
  Walsh N, Woods M. 2011  {NEST: A Comprehensive Model for Scintillation Yield
  in Liquid Xenon}. {\em JINST} \textbf{6}, P10002.
(\href{http://dx.doi.org/10.1088/1748-0221/6/10/P10002}{10.1088/1748-0221/6/10/P10002})

\bibitem{Aalbers:2022dzr}
Aalbers J et~al.. 2023  {A next-generation liquid xenon observatory for dark
  matter and neutrino physics}. {\em J. Phys. G} \textbf{50}, 013001.
(\href{http://dx.doi.org/10.1088/1361-6471/ac841a}{10.1088/1361-6471/ac841a})

\bibitem{LZ:2019sgr}
Akerib DS et~al.. 2020  {The LUX-ZEPLIN (LZ) Experiment}. {\em Nucl. Instrum.
  Meth. A} \textbf{953}, 163047.
(\href{http://dx.doi.org/10.1016/j.nima.2019.163047}{10.1016/j.nima.2019.163047})

\bibitem{PandaX-4T:2021bab}
Meng Y et~al.. 2021  {Dark Matter Search Results from the PandaX-4T
  Commissioning Run}. {\em Phys. Rev. Lett.} \textbf{127}, 261802.
(\href{http://dx.doi.org/10.1103/PhysRevLett.127.261802}{10.1103/PhysRevLett.127.261802})

\bibitem{XENON:2020kmp}
Aprile E et~al.. 2020  {Projected WIMP sensitivity of the XENONnT dark matter
  experiment}. {\em JCAP} \textbf{11}, 031.
(\href{http://dx.doi.org/10.1088/1475-7516/2020/11/031}{10.1088/1475-7516/2020/11/031})

\bibitem{LZ:2022ufs}
Aalbers J et~al.. 2023  {First Dark Matter Search Results from the LUX-ZEPLIN
  (LZ) Experiment}. {\em Phys. Rev. Lett.} \textbf{131}, 041002.
(\href{http://dx.doi.org/https://doi.org/10.1103/PhysRevLett.131.041002}{https://doi.org/10.1103/PhysRevLett.131.041002})

\bibitem{XENON:2023sxq}
Aprile E et~al.. 2023  {First Dark Matter Search with Nuclear Recoils from the
  XENONnT Experiment}. {\em Phys. Rev. Lett.} \textbf{131}, 041003.
(\href{http://dx.doi.org/10.1103/PhysRevLett.131.041003}{10.1103/PhysRevLett.131.041003})

\bibitem{Baudis:2012bc}
Baudis L. 2012  {DARWIN: dark matter WIMP search with noble liquids}. {\em J.
  Phys. Conf. Ser.} \textbf{375}, 012028.
(\href{http://dx.doi.org/10.1088/1742-6596/375/1/012028}{10.1088/1742-6596/375/1/012028})

\bibitem{Aalbers:2016jon}
Aalbers J et~al.. 2016  {DARWIN: towards the ultimate dark matter detector}.
  {\em JCAP} \textbf{1611}, 017.
(\href{http://dx.doi.org/10.1088/1475-7516/2016/11/017}{10.1088/1475-7516/2016/11/017})

\bibitem{xlzd-website}
Consortium X. 2022  {{XLZD} Dark Matter Detection Consortium}. {\url
  {https://xlzd.org}}.
Accessed: 2023-06-23.

\bibitem{Wang:2023wrr}
Wang X, Lei Z, Ju Y, Liu J, Zhou N, Chen Y, Wang Z, Cui X, Meng Y, Zhao L. 2023
   {Design, construction and commissioning of the PandaX-30T liquid xenon
  management system}. {\em JINST} \textbf{18}, P05028.
(\href{http://dx.doi.org/10.1088/1748-0221/18/05/P05028}{10.1088/1748-0221/18/05/P05028})

\bibitem{XENON:2021fkt}
Aprile E et~al.. 2022  {Application and modeling of an online distillation
  method to reduce krypton and argon in XENON1T}. {\em PTEP} \textbf{2022},
  053H01.
(\href{http://dx.doi.org/10.1093/ptep/ptac074}{10.1093/ptep/ptac074})

\bibitem{Schumann:2015cpa}
Schumann M et~al.. 2015  {Dark matter sensitivity of multi-ton liquid xenon
  detectors}. {\em JCAP} \textbf{1510}, 016.
(\href{http://dx.doi.org/10.1088/1475-7516/2015/10/016}{10.1088/1475-7516/2015/10/016})

\bibitem{Plante:2022khm}
Plante G, Aprile E, Howlett J, Zhang Y. 2022  {Liquid-phase purification for
  multi-tonne xenon detectors}. {\em Eur. Phys. J. C} \textbf{82}, 860.
(\href{http://dx.doi.org/10.1140/epjc/s10052-022-10832-w}{10.1140/epjc/s10052-022-10832-w})

\bibitem{Stifter:2020ktw}
Stifter K. 2020  {Development and performance of high voltage electrodes for
  the LZ experiment}. {\em J. Phys. Conf. Ser.} \textbf{1468}, 012016.
(\href{http://dx.doi.org/10.1088/1742-6596/1468/1/012016}{10.1088/1742-6596/1468/1/012016})

\bibitem{Baudis:2021ipf}
Baudis L, Biondi Y, Galloway M, Girard F, Manfredini A, McFadden N, Peres R,
  Sanchez-Lucas P, Thieme K. 2021  {Design and construction of Xenoscope
  \textemdash{} a full-scale vertical demonstrator for the DARWIN observatory}.
  {\em JINST} \textbf{16}, P08052.
(\href{http://dx.doi.org/10.1088/1748-0221/16/08/P08052}{10.1088/1748-0221/16/08/P08052})

\bibitem{xesat2023-website}
Julia~Mueller UoF. 2023  {Talk at XeSAT 2023}. {\url
  {https://indico.in2p3.fr/event/28661/contributions/125533/attachments/78407/114331/Pancake_Mueller_Xesat2023.pdf}}.
Accessed: 2023-07-28.

\bibitem{PandaX:2023toi}
Ning X et~al.. 2023  {Limits on the luminance of dark matter from xenon recoil
  data}. {\em Nature} \textbf{618}, 47--50.
(\href{http://dx.doi.org/10.1038/s41586-023-05982-0}{10.1038/s41586-023-05982-0})

\bibitem{OHare:2021utq}
O'Hare CAJ. 2021  {New Definition of the Neutrino Floor for Direct Dark Matter
  Searches}. {\em Phys. Rev. Lett.} \textbf{127}, 251802.
(\href{http://dx.doi.org/10.1103/PhysRevLett.127.251802}{10.1103/PhysRevLett.127.251802})

\bibitem{ParticleDataGroup:2022pth}
Workman RL et~al.. 2022  {Review of Particle Physics}. {\em PTEP}
  \textbf{2022}, 083C01.
(\href{http://dx.doi.org/10.1093/ptep/ptac097}{10.1093/ptep/ptac097})

\bibitem{XENON:2020rca}
Aprile E et~al.. 2020  {Excess electronic recoil events in XENON1T}. {\em Phys.
  Rev. D} \textbf{102}, 072004.
(\href{http://dx.doi.org/10.1103/PhysRevD.102.072004}{10.1103/PhysRevD.102.072004})

\bibitem{XENONCollaboration:2022kmb}
Aprile E et~al.. 2022  {Search for New Physics in Electronic Recoil Data from
  XENONnT}. {\em Phys. Rev. Lett.} \textbf{129}, 161805.
(\href{http://dx.doi.org/10.1103/PhysRevLett.129.161805}{10.1103/PhysRevLett.129.161805})

\bibitem{LZ:2021xov}
Akerib DS et~al.. 2021  {Projected sensitivities of the LUX-ZEPLIN experiment
  to new physics via low-energy electron recoils}. {\em Phys. Rev. D}
  \textbf{104}, 092009.
(\href{http://dx.doi.org/10.1103/PhysRevD.104.092009}{10.1103/PhysRevD.104.092009})

\bibitem{PandaX:2022xqx}
Li S et~al.. 2023  {Search for Light Dark Matter with Ionization Signals in the
  PandaX-4T Experiment}. {\em Phys. Rev. Lett.} \textbf{130}, 261001.
(\href{http://dx.doi.org/10.1103/PhysRevLett.130.261001}{10.1103/PhysRevLett.130.261001})

\bibitem{Bossio:2023wpj}
Bossio E, Agostini M. 2023  {Probing Beyond the Standard Model Physics with
  Double-beta Decays}. .

\bibitem{Wittweg:2020fak}
Wittweg C, Lenardo B, Fieguth A, Weinheimer C. 2020  {Detection prospects for
  the second-order weak decays of $^{124}$Xe in multi-tonne xenon time
  projection chambers}. {\em Eur. Phys. J. C} \textbf{80}, 1161.
(\href{http://dx.doi.org/10.1140/epjc/s10052-020-08726-w}{10.1140/epjc/s10052-020-08726-w})

\bibitem{PandaX:2022kwg}
Si L et~al.. 2022  {Determination of Double Beta Decay Half-Life of 136Xe with
  the PandaX-4T Natural Xenon Detector}. {\em Research} \textbf{2022}, 9798721.
(\href{http://dx.doi.org/10.34133/2022/9798721}{10.34133/2022/9798721})

\bibitem{LZ:2019qdm}
Akerib DS et~al.. 2020  {Projected sensitivity of the LUX-ZEPLIN experiment to
  the $0\nu\beta\beta$ decay of $^{136}Xe$}. {\em Phys. Rev. C} \textbf{102},
  014602.
(\href{http://dx.doi.org/10.1103/PhysRevC.102.014602}{10.1103/PhysRevC.102.014602})

\bibitem{LZ:2021blo}
Akerib DS et~al.. 2021  {Projected sensitivity of the LUX-ZEPLIN experiment to
  the two-neutrino and neutrinoless double $\beta$ decays of $^{134}$Xe}. {\em
  Phys. Rev. C} \textbf{104}, 065501.
(\href{http://dx.doi.org/10.1103/PhysRevC.104.065501}{10.1103/PhysRevC.104.065501})

\bibitem{KamLAND-Zen:2022tow}
Abe S et~al.. 2023  {Search for the Majorana Nature of Neutrinos in the
  Inverted Mass Ordering Region with KamLAND-Zen}. {\em Phys. Rev. Lett.}
  \textbf{130}, 051801.
(\href{http://dx.doi.org/10.1103/PhysRevLett.130.051801}{10.1103/PhysRevLett.130.051801})

\bibitem{nEXO:2021ujk}
Adhikari G et~al.. 2022  {nEXO: neutrinoless double beta decay search beyond
  10$^{28}$ year half-life sensitivity}. {\em J. Phys. G} \textbf{49}, 015104.
(\href{http://dx.doi.org/10.1088/1361-6471/ac3631}{10.1088/1361-6471/ac3631})

\bibitem{Agostini:2020adk}
Agostini F et~al.. 2020  {Sensitivity of the DARWIN observatory to the
  neutrinoless double beta decay of $^{136}$Xe}. {\em Eur. Phys. J. C}
  \textbf{80}, 808.
(\href{http://dx.doi.org/10.1140/epjc/s10052-020-8196-z}{10.1140/epjc/s10052-020-8196-z})

\bibitem{DARWIN:2020bnc}
Aalbers J et~al.. 2020  {Solar neutrino detection sensitivity in DARWIN via
  electron scattering}. {\em Eur. Phys. J. C} \textbf{80}, 1133.
(\href{http://dx.doi.org/10.1140/epjc/s10052-020-08602-7}{10.1140/epjc/s10052-020-08602-7})

\bibitem{Freedman:1973yd}
Freedman DZ. 1974  {Coherent Neutrino Nucleus Scattering as a Probe of the Weak
  Neutral Current}. {\em Phys. Rev. D} \textbf{9}, 1389--1392.
(\href{http://dx.doi.org/10.1103/PhysRevD.9.1389}{10.1103/PhysRevD.9.1389})

\bibitem{GargamelleNeutrino:1973jyy}
Hasert FJ et~al.. 1973  {Observation of Neutrino Like Interactions Without Muon
  Or Electron in the Gargamelle Neutrino Experiment}. {\em Phys. Lett. B}
  \textbf{46}, 138--140.
(\href{http://dx.doi.org/10.1016/0370-2693(73)90499-1}{10.1016/0370-2693(73)90499-1})

\bibitem{COHERENT:2017ipa}
Akimov D et~al.. 2017  {Observation of Coherent Elastic Neutrino-Nucleus
  Scattering}. {\em Science} \textbf{357}, 1123--1126.
(\href{http://dx.doi.org/10.1126/science.aao0990}{10.1126/science.aao0990})

\bibitem{COHERENT:2020iec}
Akimov D et~al.. 2021  {First Measurement of Coherent Elastic Neutrino-Nucleus
  Scattering on Argon}. {\em Phys. Rev. Lett.} \textbf{126}, 012002.
(\href{http://dx.doi.org/10.1103/PhysRevLett.126.012002}{10.1103/PhysRevLett.126.012002})

\bibitem{Xiang:2023csc}
Xiang X et~al.. 2023  {Nuclear recoil response of liquid xenon and its impact
  on solar 8B neutrino and dark matter searches}. .

\bibitem{XENON:2020gfr}
Aprile E et~al.. 2021  {Search for Coherent Elastic Scattering of Solar $^8$B
  Neutrinos in the XENON1T Dark Matter Experiment}. {\em Phys. Rev. Lett.}
  \textbf{126}, 091301.
(\href{http://dx.doi.org/10.1103/PhysRevLett.126.091301}{10.1103/PhysRevLett.126.091301})

\bibitem{PandaX:2022aac}
Ma W et~al.. 2023  {Search for Solar B8 Neutrinos in the PandaX-4T Experiment
  Using Neutrino-Nucleus Coherent Scattering}. {\em Phys. Rev. Lett.}
  \textbf{130}, 021802.
(\href{http://dx.doi.org/10.1103/PhysRevLett.130.021802}{10.1103/PhysRevLett.130.021802})

\bibitem{Baudis:2013qla}
Baudis L et~al.. 2014  {Neutrino physics with multi-ton scale liquid xenon
  detectors}. {\em JCAP} \textbf{01}, 044.
(\href{http://dx.doi.org/10.1088/1475-7516/2014/01/044}{10.1088/1475-7516/2014/01/044})

\bibitem{Lang:2016zhv}
Lang RF, McCabe C, Reichard S, Selvi M, Tamborra I. 2016  {Supernova neutrino
  physics with xenon dark matter detectors: A timely perspective}. {\em Phys.
  Rev.} \textbf{D94}, 103009.
(\href{http://dx.doi.org/10.1103/PhysRevD.94.103009}{10.1103/PhysRevD.94.103009})

\bibitem{PeresRicardoThesis2023}
Mota~Peres RJ. 2023 {\em Advancing Multi-Messenger Astrophysics and Dark Matter
  Searches with XENONnT and the Top SiPM Array of Xenoscope}.
PhD thesis University of Zurich, 2023.

\bibitem{snews-website}
Network S. 2023  {{SNEWS} Supernova Early Warning System}. {\url
  {https://snews.bnl.gov}}.
Accessed: 2023-07-22.

\bibitem{Suliga:2021hek}
Suliga AM, Beacom JF, Tamborra I. 2022  {Towards probing the diffuse supernova
  neutrino background in all flavors}. {\em Phys. Rev. D} \textbf{105}, 043008.
(\href{http://dx.doi.org/10.1103/PhysRevD.105.043008}{10.1103/PhysRevD.105.043008})

\bibitem{Baudis:2012ig}
Baudis L. 2012  {Direct dark matter detection: the next decade}. {\em Phys.
  Dark Univ.} \textbf{1}, 94--108.
(\href{http://dx.doi.org/10.1016/j.dark.2012.10.006}{10.1016/j.dark.2012.10.006})

\end{thebibliography}

\end{document}